\documentclass[twocolumn,tighten,floatfix]{aastex631} 
\pdfoutput=1 
\usepackage{amsmath,amstext}
\usepackage[T1]{fontenc}
\usepackage[figure,figure*]{hypcap}
\usepackage{threeparttable}
\usepackage[htt]{hyphenat}
\usepackage{float}
\usepackage{orcidlink}
\usepackage{bm}
\usepackage{bbold}
\usepackage{xspace}
\usepackage{xcolor}
\usepackage[caption=false]{subfig}
\usepackage{graphicx}
\usepackage{afterpage}
\graphicspath{{images/}}

\hypersetup{colorlinks=true,
            citecolor=[rgb]{0, .4, .8},
            linkcolor=[rgb]{0, .4, .8},
            urlcolor=[rgb]{0, .4, .8}}

\newcommand{\aemulus}{{\sc Aemulus}\xspace}

\newcommand{\inv}{^{-1}}
\newcommand{\T}{^\top}
\newcommand{\like}{\mathcal{L}}
\newcommand{\cov}[1]{C_\text{#1}}
\newcommand{\covtot}{C_\like}
\newcommand{\simo}{\mathord{\sim}}

\newcommand{\hMpc}{h\inv\,\mathrm{Mpc}\xspace}
\newcommand{\hGpc}{h\inv\,\mathrm{Gpc}\xspace}
\newcommand{\Mpch}{h\,\mathrm{Mpc}\inv\xspace}

\newcommand{\wprp}{w_\mathrm{p}(r_\mathrm{p})\xspace}
\newcommand{\cfm}{\xi_0(s)\xspace}
\newcommand{\cfq}{\xi_2(s)\xspace}
\newcommand{\upf}{P_\mathrm{U}(s)\xspace}
\newcommand{\mcf}{M(s)\xspace}

\newcommand{\om}{\Omega_\mathrm{m}\xspace}
\newcommand{\sig}{\sigma_\mathrm{8}\xspace}
\newcommand{\gf}{\gamma_f\xspace}
\newcommand{\gfs}{\gamma_f f\sigma_8\xspace}
\newcommand{\fsig}{f\sigma_8 \xspace}
\newcommand{\msat}{M_\mathrm{sat}\xspace}
\newcommand{\mcut}{M_\mathrm{cut}\xspace}
\newcommand{\vbs}{v_\mathrm{bs}\xspace}
\newcommand{\fenv}{f_\mathrm{env}\xspace}
\newcommand{\fmax}{f_\mathrm{max}\xspace}

\newcommand{\new}[1]{#1}

\begin{document}

\shorttitle{Aemulus VI: Beyond-standard statistics}

\title{The Aemulus Project VI: Emulation of beyond-standard galaxy clustering statistics to improve cosmological constraints}

\author[0000-0001-8764-7103]{Kate Storey-Fisher}
\affiliation{Center for Cosmology and Particle Physics, Department of Physics, New York University, 726 Broadway, New York, NY 10003, USA}

\author[0000-0003-3578-6149]{Jeremy L. Tinker}
\affiliation{Center for Cosmology and Particle Physics, Department of Physics, New York University, 726 Broadway, New York, NY 10003, USA}
 
\author[0000-0001-7984-5476]{Zhongxu Zhai}
\affiliation{Department of Astronomy, School of Physics and Astronomy, Shanghai Jiao Tong University, Shanghai 200240, China}
\affiliation{Shanghai Key Laboratory for Particle Physics and Cosmology, Shanghai 200240, China}
\affiliation{Waterloo Center for Astrophysics, University of Waterloo, Waterloo, ON N2L 3G1, Canada}
\affiliation{Department of Physics and Astronomy, University of Waterloo, Waterloo, ON N2L 3G1, Canada}

\author[0000-0002-0728-0960]{Joseph DeRose} 
\affiliation{Lawrence Berkeley National Laboratory, 1 Cyclotron Road, Berkeley, CA 94720, USA}

\author[0000-0003-2229-011X]{Risa H. Wechsler}
\affiliation{Kavli Institute for Particle Astrophysics and Cosmology and Department of Physics, Stanford University, Stanford, CA 94305, USA}
\affiliation{SLAC National Accelerator Laboratory, Menlo Park, CA 94025, USA}
\affiliation{Department of Physics, Stanford University, 382 Via Pueblo Mall, Stanford, CA 94305, USA}

\author[0000-0002-5209-1173]{Arka Banerjee}
\affiliation{Department of Physics, Indian Institute of Science Education and Research, Homi Bhabha Road, Pashan, Pune 411008, India}

\correspondingauthor{Kate Storey-Fisher}
\email{k.sf@nyu.edu}

\begin{abstract}
There is untapped cosmological information in galaxy redshift surveys in the nonlinear regime.
In this work, we use the \aemulus suite of cosmological $N$-body simulations to construct Gaussian process emulators of galaxy clustering statistics at small scales ($0.1$--$50 \: \hMpc$) in order to constrain cosmological and galaxy bias parameters.
In addition to standard statistics---the projected correlation function $\wprp$, the redshift-space monopole of the correlation function $\cfm$, and the quadrupole $\cfq$---we emulate statistics that include information about the local environment, namely the underdensity probability function $\upf$ and the density-marked correlation function $\mcf$.
This extends the model of \aemulus III for redshift-space distortions by including new statistics sensitive to galaxy assembly bias.
In recovery tests, we find that the beyond-standard statistics significantly increase the constraining power on cosmological parameters of interest: including $\upf$ and $\mcf$ improves the precision of our constraints on $\om$ by 27\%, $\sig$ by 19\%,  and the growth of structure parameter, $\fsig$, by 12\% compared to standard statistics.
We additionally find that scales below $\simo6 \: \hMpc$ contain as much information as larger scales.
The density-sensitive statistics also contribute to constraining halo occupation distribution parameters and a flexible environment-dependent assembly bias model, which is important for extracting the small-scale cosmological information as well as understanding the galaxy--halo connection.
This analysis demonstrates the potential of emulating beyond-standard clustering statistics at small scales to constrain the growth of structure as a test of cosmic acceleration.
\end{abstract}

\keywords{\href{http://astrothesaurus.org/uat/902}{Large-scale structure of the universe (902)}, \href{http://astrothesaurus.org/uat/339}{Cosmological parameters(339)}, \href{http://astrothesaurus.org/uat/1965}{Computational methods (1965)}, \href{http://astrothesaurus.org/uat/1882}{Astrostatistics (1882)}}

\section{Introduction}

Galaxy redshift surveys contain a wealth of information about the cosmological model.
Galaxies trace the underlying matter distribution, and their clustering gives us detailed insight into the growth history of the universe.
Recent spectroscopic surveys, including SDSS \citep{York2000} and its extensions BOSS \citep{Dawson2013} and eBOSS \citep{Dawson2015}, have provided impressive constraints on cosmology using galaxy clustering.
Upcoming surveys such as DESI \citep{Aghamousa2016}, the Subaru Prime Focus Spectrograph \citep{takada_extragalactic_2014}, and eventually Euclid \citep{Laureijs2011} and the Nancy Grace Roman Space Telescope \citep{Green2012}, will measure tens of millions of spectroscopic redshifts, allowing for unprecedented cosmological measurements.

Most of the current state-of-the-art constraints from these data sets are based on galaxy clustering at large scales.
One of the main probes used to measure the growth of structure in spectroscopic analyses is redshift-space distortions (RSDs), anisotropies in clustering induced by galaxy peculiar velocities.
For the scales over which the RSD effect is typically analyzed, around $\simo40$--$150 \, \hMpc$, the evolution of matter is close to linear and can be modeled with linear perturbation theory (e.g., \citealt{Alam2017}).
While this approach has been very successful, current and future surveys will be most precise at much smaller scales, given their requirements on galaxy number density.
It is not currently known how much additional information exists at these small scales, but recent work suggests that it is significant and may even exceed the information content at large scales \citep{Zhai2019}.
Extracting this information requires accurately modeling the nonlinear dynamics of dark matter down to these scales.
Cosmological $N$-body simulation have been remarkably successful at this (e.g., \citealt{Klypin2011}); however, they are very expensive to run, and including hydrodynamics is intractable for complete cosmological inference purposes.

In order to use $N$-body simulations for cosmological analysis, we require a galaxy bias model to populate the dark matter distribution with galaxies \citep{Seljak2000, BerlindWeinberg2002, CooraySheth2002, Zheng2005}, which probabilistically describes the occupation number of galaxies in dark matter halos as a function halo mass.
The simple HOD model reconstructs galaxy clustering to a reasonable degree of accuracy; however, it has been shown that occupation has a small but non-negligible dependence on secondary halo properties, known as galaxy assembly bias (see, e.g., \citealt{Wechsler2006, Croton2007, Zentner2014, WechslerTinker2018}).
Modeling assembly bias is critical for obtaining the most accurate cosmological constraints, as well as understanding the galaxy--halo connection.

Late-time galaxy clustering analyses have put increasingly strong constraints on the growth of structure parameter $f \sigma_8$.
While some of these agree with results from the cosmic microwave background as measured by Planck \citep{alam_completed_2021,zhang_boss_2022}, others are in $1\sigma$--$4\sigma$ tension (e.g., \citealt{Macaulay2013, Sanchez2014, deMattia2021}).
A series of recent studies focusing on small scales have also found a few-sigma tension \citep{Chapman2021, Lange2022, Yuan2022, zhai_aemulus_2023}, and these agree with the results of weak-lensing studies (e.g., \citealt{MacCrann2015, Leauthaud2017, joudaki_kidsviking-450_2020}).
Improving the constraining power from clustering analyses is important for determining if the tension still holds; one avenue for doing this is expanding beyond RSD to include other clustering statistics.

Current cosmological analyses focus on a small set of two-point statistics of galaxy clustering that are well-understood theoretically.
While these statistics are highly informative, it has been shown that there is significant additional information in other nonstandard observables.
For instance, \cite{Tinker2006, Tinker2008} demonstrated that the void probability function and underdensity probability function contribute complementary information to two-point statistics, due to their sensitivity to the environmental dependence of halo occupation.
Other work has demonstrated the constraining power in these and other related counts-in-cells statistics \citep{WalshTinker2019, Wang2019, Beltz-Mohrmann2020}.

The marked correlation function \citep{Sheth2004} has also been shown to contain information complementary to that in standard statistics.
\cite{WhitePadmanabhan2009} demonstrated that when using a local density-based mark, the statistic is useful in constraining the cosmological parameter $\sig$ by breaking degeneracies in HOD modeling; \cite{White2016} found that it is sensitive to modifications to general relativity.
Recently, \cite{Szewciw2022} aimed to optimally constrain the galaxy--halo connection, and confirmed that including the marked correlation function, as well as counts-in-cells statistics and others including the group multiplicity function and group velocity dispersion, significantly improve constraints on halo model parameters at fixed cosmology.

In this work, we combine the use of beyond-standard clustering statistics with the emulation approach.
Emulation has recently been explored as a method for making highly accurate predictions for cosmology at nonlinear scales while minimizing requirements on cosmological simulations \citep{Heitmann2009, Heitmann2010, Lawrence2010}.
The idea is to first construct a sparse training set of high-resolution $N$-body simulations that span the allowable parameter space.
Then a model can be trained to make fast predictions of the output of the simulations, or summary statistics of the output, given the input parameters.
This can finally be used in inference to fully explore the parameter space, essentially interpolating in high dimensions over the regions between input simulations.
Machine learning models are often used for this purpose, due to the need to model such a high-dimensional space and produce quick predictions.

Cosmological emulators typically aim to predict summary statistics of the matter and galaxy distributions.
Two-point statistics, namely the power spectrum and its real-space counterpart the correlation function, are the key observables used to constrain cosmological models.
There has been significant work emulating the matter power spectrum \citep{Heitmann2009, Lawrence2017, Giblin2019, Ho2022}.
Recent work has extended and improved upon this approach, such as the incorporation of dynamical dark energy and massive neutrinos into emulators \citep{Angulo2021}, and the development of fully differentiable power spectrum emulators \citep{SpurioMancini2022, derose_neural_2022}.
Other emulators predict the galaxy power spectrum \citep{Kwan2015, Pellejero-Ibanez2020, kokron_cosmology_2021}, and \cite{Wibking2019} recently emulated the galaxy correlation function along with galaxy--galaxy lensing.

Simulation-based emulators have been used to improve precision on cosmological parameter constraints from recent surveys:
\cite{Miyatake2021} constrain $S_8$ from the HSC-Y1 and SDSS data using the \textsc{DarkEmulator} \citep{Nishimichi2019}.
\cite{neveux_combined_2022} apply a Gaussian process emulator to the BOSS galaxy and eBOSS quasar samples, obtaining constraints similar to those of SDSS using half the amount of data.
\cite{EuclidPrepII} constructed the \textsc{EuclidEmulator} to predict the nonlinear correction of the matter power spectrum in preparation for the upcoming Euclid survey; the improved version \citep{EuclidPrepIX} achieves 1\% accuracy or better for $0.01 \: \Mpch \leq k \leq 10 \: \Mpch$.

This work is part of the \aemulus Project, which uses a suite of high-resolution $N$-body simulations expressly designed for emulation at small scales to improve cosmological constraints.
The previous papers in the project introduce the simulation suite \citep{DeRose2018} and construct emulators of the halo mass function \citep{McClintock2018}, the galaxy correlation function \citep{Zhai2019}, and halo bias \citep{McClintock2019}.
The \aemulus emulator has been used to constrain the growth rate of structure in the BOSS-LOWZ sample \citep{Lange2022} and the eBOSS LRG sample \citep{Chapman2021}, both obtaining nearly a factor-of-two increase in precision on $\fsig$ compared to standard measurements at linear scales.
Most recently, the \aemulus project constructed two-point function emulators that include models of assembly bias and deviations from general relativity (GR) to provide improved precision on the growth rate of structure parameter from the BOSS survey \citep{zhai_aemulus_2023}.

In this paper, we extend the work of \aemulus III \citep{Zhai2019} to include emulation of two beyond-standard observables:
The underdensity probability function $\upf$, defined as the probability that a randomly placed sphere has a galaxy density less than some threshold (e.g., \citealt{HoyleVogeley2004}), and the marked correlation function $\mcf$, the two-point correlation function with galaxy pairs weighted by their properties \citep{Sheth2004}.
We extend the HOD model of \aemulus III to include a model of assembly bias, based on the local density.
We also incorporate several more HOD parameters for increased flexibility, as well as a parameter that scales \new{the} velocity field to model deviations from GR, following \aemulus V.

This paper is organized as follows:
In \S\ref{sec:sims_gals}, we describe the $N$-body simulations and halo occupation distribution model used, and in \S\ref{sec:observables}, we outline the five clustering statistics we use for inference.
We detail our emulation and inference methods in \S\ref{sec:methods}, and show the results of recovery tests on both \aemulus mocks and an external mock catalog in \S\ref{sec:results}.
In \S\ref{sec:discussion}, we discuss the implications of these results and our conclusions.

\section{Simulations and Galaxy Bias Model}
\label{sec:sims_gals}

In this section, we detail the \aemulus $N$-body simulations that are used as the basis for our emulation (\S\ref{sec:aemulus}), and the halo occupation distribution model used to model the galaxy--halo connection and populate the simulations to construct mock galaxy catalogs (\S\ref{sec:hod}). 

\subsection{The Aemulus simulations}
\label{sec:aemulus}

We use the \aemulus simulations, a suite of 75 high-resolution $N$-body simulations \citep{DeRose2018}.
They have a box size $L = 1.05$ $\hGpc$ with $1400^3$ dark matter particles, and a mass resolution of $\simo3.5\times10^{10}h\inv \: M_{\odot}$ (depending on the cosmology).
The training set consists of 40 different $w$CDM cosmologies, selected using a Latin hypercube to span the parameter space.
\new{The ranges of the cosmological parameters are shown in Table 3 of \aemulus V \citep{zhai_aemulus_2023}.}
The test set is comprised of seven different cosmologies, with five realizations with different initial conditions for each cosmology, totaling 35 test boxes \new{(with a slightly reduced parameter space compared to the training simulations)}.
We use the redshift $z=0.55$ snapshot for this work.
We use the training set to train our emulator, and the test set to verify its performance as well as to estimate the sample variance.

Our cosmological model consists of seven parameters: the matter energy density $\om$, the baryon energy density $\Omega_{b}$, the amplitude of matter fluctuations $\sig$, the dimensionless Hubble constant $h$, the spectral index of the primordial power spectrum $n_{s}$, the dark energy equation of state parameter $w$, and the number of relativistic species $N_{\text{eff}}$.
These simulations are based on GR, so we include a scaling parameter $\gamma_f$ to capture non-GR effects; it is defined as the amplitude of the halo velocity field relative to the $w$CDM+GR prediction.
The parameters of interest for this work are $\om$, $\sig$, and $\gamma_f$; we do not expect our approach to be particularly sensitive to the other parameters \citep{Zhai2019}, and these are marginalized over.
Most importantly, we are interested in the growth of structure parameter $f \sigma_8$, and we parameterize it to be independent of GR by including the velocity field scaling parameter $\gf$ \citep{Reid2014}.
We henceforth compute and refer to the growth of structure parameter as $\gfs$.

\subsection{Halo occupation distribution model}
\label{sec:hod}

To create mock galaxy catalogs from these simulations, we use the halo occupation distribution to model the galaxy--halo connection.
The HOD framework starts from the assumption that the number of galaxies $N$ in a given dark matter halo depends only on the mass of the host halo $M$, and gives a probability distribution for $N$ given $M$: $P(N|M)$.  
We base our HOD model on those of \cite{Zheng2005} and \cite{reddick_connection_2013}, which separate the contribution of central and satellite galaxies, $\langle N(M) \rangle = \langle N_\mathrm{cen}(M) \rangle + \langle N_\mathrm{sat}(M) \rangle$.
The central galaxy occupation function is modeled as a Bernoulli distribution with a mean of
\begin{equation}
	\langle N_\mathrm{cen}(M) \rangle = \frac{\fmax}{2} \left[ 1 + \mathrm{erf}
	\left(\frac{\mathrm{log}_{10} M - \mathrm{log}_{10} M_\mathrm{min} }{\sigma_{\mathrm{log}M}}\right) \right] ~,
\end{equation}
where \texttt{erf()} is the error function. 
The number of satellite galaxies is drawn from a Poisson distribution with a mean of 
\begin{equation}
	\langle N_\mathrm{sat}(M) \rangle = \left( \frac{M}{\msat} \right)^\alpha
	\mathrm{exp} \left( - \frac{M_\mathrm{cut}}{M} \right) N_\mathrm{cen}(M) ~.
\end{equation}
The parameters are defined as follows: $M_\mathrm{min}$ is the mass at which half of the halos host a central galaxy, $\sigma_{\mathrm{log}M}$ controls the scatter of halo mass at fixed galaxy luminosity, $\alpha$ is the power-law index for the mass dependence of the number of satellites, $\msat$ is a typical mass for halos to host one satellite, $M_\mathrm{cut}$ varies the cutoff mass in the satellite occupation function, \new{and $\fmax$ is the central occupation fraction of the highest-mass halos.
When the $\fmax$ parameter equals unity, in the high halo mass limit, all halos host galaxies; setting $\fmax < 1$ adjusts this fraction. 
This accounts for bright galaxies missed in target selection---for example, due to color and magnitude effects, as is the case in the BOSS-LOWZ sample \citep{leauthaud_stripe_2016}; a similar parameterization has been used in other analyses \citep{hoshino_luminous_2015, guo_incomplete_2018, Chapman2021, zhai_aemulus_2023}.
}

We fix the number density to $\bar{n} = 2 \times 10^{-4} \, (\hMpc)^{-3}$ \new{by computing} $M_\mathrm{min}$ to satisfy this number density after varying the other HOD parameters.
This value is somewhat lower than the peak BOSS number density, but similar to that of a luminous red galaxy (LRG) sample, and it is designed to produce a sample closer to being volume limited; it is the number density used in the \aemulus V analysis of BOSS-LOWZ+CMASS \citep{zhai_aemulus_2023}.
\new{We note that the amplitude of density fluctuations is degenerate with the galaxy bias to linear order, so fixing the number density risks artificially breaking this degeneracy and biasing the results.
However, our inclusion of the $\fmax$ parameter effectively allows for flexibility in the galaxy bias, as it sets a ceiling for the central galaxy occupation of the halo field.
This has been shown by \cite{Chapman2021} (Section 4.2): with a fixed number density emulator, they demonstrate that fixing $\fmax=1$ results in a bias in the recovered halo velocity field rescaling parameter $\gf$, while freeing $\fmax$ eliminates this bias.
\cite{Chapman2021} also perform a test of the Alcock--Paczynski scaling effect \citep{AlcockPaczynski1979} (Section 3.6), which impacts the number density, and find that this change has a negligible effect on final constraints.
Additionally, in our target sample, BOSS CMASS and LOWZ, the number density is well measured: we estimate the variation in number density using the BOSS QPM mocks, and find that the fractional uncertainty is 0.43\%.
To test that this small uncertainty does not affect our results, we construct mocks with 1\% greater and lower number density than that at which the emulator is constructed; we find that the recovered parameters shift by only $\simo0.25\sigma$, and none more than $1\sigma$.
We thus conclude that fixing the number density while having a free $\fmax$ in our emulator should allow for unbiased inference of cosmology.}

We include three additional parameters \new{in our HOD model} related to halo occupation, following \cite{Zhai2019}.
In addition to the parameter $\gf$ described in \S\ref{sec:aemulus} that rescales all halo velocities, we include velocity bias parameters for galaxies relative to the virial velocity of their DM halo $\sigma_\mathrm{halo}$.
We define $v_\mathrm{bc}$ as the velocity bias of central galaxies, which rescales the velocity of centrals $\sigma_\mathrm{cen}$ relative to that of host halos as $\sigma_\mathrm{cen} = v_\mathrm{bc} \,\sigma_\mathrm{halo}$.
The velocity bias of satellite galaxies $v_\mathrm{bs}$ is defined in the same way as $v_\mathrm{bc}$.
We also include a concentration parameter relating satellite and halo concentrations, where the concentration $c$ is defined as the ratio between the halo outer radius and the scale radius (which depends on the halo density profile).
We define the concentration ratio $c_\mathrm{vir}$ as the ratio between the concentration of satellites and DM halos, $c_\mathrm{vir} = c_\mathrm{sat}/c_\mathrm{halo}$.

We extend this standard HOD model to take into account the dependence on properties other than just the host halo mass; this secondary dependence is known as assembly bias.
Here, we use the three-parameter assembly bias model of \cite{WalshTinker2019}, which includes a dependence on the local dark matter density around a halo, because we might expect the external environment of halos to play a role in galaxy formation.
Specifically, we define $\delta$ as the relative density in a sphere of radius 10 $\hMpc$ around a halo center.
The assembly bias model adjusts the minimum halo mass needed to host a central galaxy, $M_\mathrm{min}$, to a threshold $M_\mathrm{min}'$ based on the local density.
It is defined as
\begin{equation}
	M_\mathrm{min}' = M_\mathrm{min} \left[ 1 + \fenv \, \mathrm{erf}
	\left(\frac{ \delta - \delta_\mathrm{env} }{\sigma_\mathrm{env}}\right) \right] ~,
\end{equation}
where $\fenv$ controls the strength of the environmental dependence, $\delta_\mathrm{env}$ is the density threshold at which to move around satellites, and $\sigma_\mathrm{env}$ controls the sharpness of the transition between overdense and underdense regions.
A value of $\fenv > 0$ means that a halo in a higher-density environment requires a higher mass to host a central galaxy, and a halo in a lower-density environment needs a lower mass, effectively moving galaxies from high- to low-density regions.
Conversely, $\fenv < 0$ moves galaxies from low- to high-density regions.
Setting $\fenv = 0$ turns off assembly bias.

\new{After using the HOD to populate the simulation boxes with galaxies, we input redshift-space distortions. 
We do this by} projecting the real-space positions along one of the axes $x_\mathrm{r}$ into redshift-space positions $x_\mathrm{s}$:
\begin{equation}
    x_\mathrm{s} = x_\mathrm{r} + (1+z)\frac{v}{H(z)} ~,
\end{equation}
where $z$ is the redshift of the simulation, $v$ is the velocity of the galaxy along axis $x_\mathrm{r}$, and $H(z)$ is the Hubble parameter at that redshift for the given cosmology.

\new{The HOD parameter ranges we use are the same as in \aemulus V \citep{zhai_aemulus_2023}, shown in Table 3 of that work.}
We populate each of the 40 training boxes with 100 unique HOD models\new{, chosen using the Latin hypercube method \citep{Heitmann2009} with a total of 4000 samples}.
We populate the test boxes with another independent set of 100 HOD models\new{, from a separate 100 sample draw from a Latin hypercube (it should be noted that,} for the test set, we use the same 100 models to populate each of the 35 boxes, while for the training set every model is different).
This results in a training set of 4000 catalogs and a test set of 3500 catalogs for the emulator. 
(We found that two of the 4000 training mocks resulted in unphysical values of clustering statistics and discarded these from our training set.)
For the recovery tests, we use a subset of this test set consisting of 70 catalogs, with 10 unique HOD models per cosmology \new{(complete recovery tests on all 3500 models would be both expensive and repetitive, but we do use the additional models for select tests as well as for sample variance estimation)}.
Our complete model has seven cosmology parameters plus $\gf$, seven HOD parameters, and three assembly bias parameters, for a total of 18 free parameters.
These are the parameters that will be the inputs to our emulators and that we will later infer through Markov Chain Monte Carlo, based on the measured observables.

\section{Observables}
\label{sec:observables}

The goal of this work is to investigate the information in small-scale clustering using both standard statistics and other, beyond-standard observables that may contain important information.
(It should be noted that we use the words ``observables'' and ``statistics'' interchangeably in this work.)
The standard observables we use are:
\begin{itemize}
    \item The projected correlation function, $\wprp$ (\S\ref{sec:wprp});
    \item The monopole of the two-point correlation function, $\cfm$ (\S\ref{sec:cfs}); and
    \item The quadrupole of the two-point correlation function, $\cfq$ (\S\ref{sec:cfs}).
\end{itemize}
The beyond-standard observables we include are:
\begin{itemize}
    \item The underdensity probability function, $\upf$ (\S\ref{sec:upf}); and
    \item The marked correlation function, $\mcf$ (\S\ref{sec:mcf}).
\end{itemize}
We discuss the covariances between these statistics in \S\ref{sec:cov}.
The statistics measured in the given bins are shown in Figure~\ref{fig:emu_accuracy} (circles in top panel), for the 3500 test set models.

\subsection{The projected correlation function, \texorpdfstring{$\wprp$}{wp(rp)}}
\label{sec:wprp}

The two-point correlation function is defined as the excess probability above a Poisson random distribution that two galaxies are separated by a given distance $r$.
In practice, we work in redshift space with vector distance $\bm{s}$, defining  $\bm{s} = \bm{s}_2 -  \bm{s}_1$ and  $\bm{l} = ( \bm{s}_1 + \bm{s}_2)/2$.
We measure the two-dimensional correlation function $\xi_\mathrm{Z}(r_\mathrm{p}, \pi)$ on a grid, where the subscript Z denotes redshift-space, $\pi$ is the transverse separation, and $r_\mathrm{p}$ is the line-of-sight separation, defined as
\begin{equation}
	\pi = \frac{\bm{s}\cdot\bm{l}}{|\bm{l}|}, 
	\quad r_{p}^2 =\bm{s}\cdot\bm{s}-\pi^2 ~.
\end{equation}
Then, the projected correlation function is
\begin{equation}
	\wprp = 2 \int_{0}^{\infty} d\pi \, \xi_\mathrm{Z}(r_\mathrm{p}, \pi) ~.
\end{equation}
In practice, we cut off the integral at a scale of $\pi_\mathrm{max} = 40 \, \hMpc$.
This choice of a somewhat low $\pi_\mathrm{max}$ leaves $\wprp$ sensitive to RSDs in the two-halo term.
However, this preserves some cosmological information, and in any case, it is consistent in the constructed emulator, so it will not lead to a bias in parameter recovery.

We must use an estimator to measure the correlation function in data.
We use the natural estimator \citep{PeeblesHauser1974},
\begin{equation}
	\xi(r_\mathrm{p}, \pi) = \frac{DD}{RR} - 1,
\end{equation}
where DD is the number of data--data pairs in an $(r_\mathrm{p}, \pi)$ bin, and RR is the number of random--random pairs in a uniform random catalog of the same size as the data, each normalized by the total number of galaxy pairs in the respective catalog pair.
Because we are working with periodic simulation boxes in this analysis, we can analytically compute the random--random (RR) term and only have to numerically compute the DD term.
We note that the DR term cannot be analytically computed, so to avoid needing a random catalog, we do not use a lower--variance estimator such as the standard \cite{LandySzalay1993} estimator.
As there is no complex window function to introduce biases or additional noise, and we are doing inference using simulations rather than model comparison, the natural estimator should be sufficient.

We measure $\wprp$ in nine logarithmically spaced bins between $r_p = 0.1$ and $50 \, \hMpc$.
We use the software package \texttt{corrfunc} \citep{SinhaGarrison2019, Sinha2020} to compute this observable.

\subsection{The two-point correlation function multipoles, \texorpdfstring{$\cfm$ and $\cfq$}{xi0(s) and xi2(s)}} 
\label{sec:cfs}

We also measure the multipoles of the redshift-space correlation, now defining the coordinates $s = |\bm{\mathrm{s}}|$ and $\mu = r_\mathrm{p}/s$:
\begin{equation}
\xi_{\ell}(s) = \frac{2\ell+1}{2}\int_{-1}^{1} L_{\ell}(\mu) \, \xi_\mathrm{Z}(s, \mu) \, d\mu ~,
\end{equation}
where $L_{\ell}$ is the Legendre polynomial of order $\ell$ (and $\ell$ indexes the multipole).
Most of the information is contained in the few lowest-order multipoles, so for this analysis, we use only the monopole $\cfm$ and the quadrupole $\cfq$.
We use the \cite{PeeblesHauser1974} estimator as in the previous section to measure the correlation functions in practice.

For $\cfm$ and $\cfq$, we use the same nine bins as we did for $\wprp$, between $s = 0.1$ and $50 \, \hMpc$, and we use 15 $\mu$ bins.
We use \texttt{corrfunc} \citep{SinhaGarrison2019, Sinha2020} and \texttt{halotools} \citep{Hearin2017} to compute these statistics.

\subsection{The underdensity probability function, \texorpdfstring{$\upf$}{P(s)}}
\label{sec:upf}

The first beyond-standard statistic we use in our analysis is the underdensity probability function, $\upf$ (e.g., \citealt{HoyleVogeley2004}).
$\upf$ is defined as the fraction of randomly placed spheres that are underdense compared to some threshold density.
This is a more robust metric to measure than the void probability function, which uses a threshold of zero and is more sensitive to issues such as the angular selection function, shot noise, and fiber collisions.
We can write $\upf$ as
\begin{equation}
	\upf = \frac{1}{N} \sum_i^N \mathbb{1}(n_i(s) < n_\mathrm{thresh}) ~,
\end{equation}
where $i$ indexes the $N$ spheres, $n_i(s)$ is the number density of galaxies in sphere $i$ with radius $s$, $\mathbb{1}()$ is an indicator function that is 1 if its argument is true and 0 otherwise, and $n_\mathrm{thresh}$ is the threshold number density. 
We choose $N=10^6$ and $n_\mathrm{thresh} = 0.2 \bar{n}$, where $\bar{n}$ is the mean number density of the mock; this is the same value chosen by \cite{HoyleVogeley2004}, which is slightly denser than the mean underdensity of large voids in the 2dF Galaxy Redshift Survey \citep{Colless2003}.

The $\upf$ does not vary significantly at small scales ($s \lesssim 5 \hMpc$) across different cosmology and HOD models (see the sample variance at small scales in Figure~\ref{fig:emu_accuracy}), so these scales are not as useful for parameter inference.
Thus, we use nine linearly spaced radii between $s = 5$ and $45 \, \hMpc$.
To compute the statistic, we modify a standard $k$-d tree code\footnote{\url{https://github.com/jtsiomb/kdtree}} to work on a periodic box.\footnote{\url{https://github.com/kstoreyf/clust}} 

\subsection{The marked correlation function, \texorpdfstring{$\mcf$}{M(s)}}
\label{sec:mcf}

The other beyond-standard statistic we investigate is the marked correlation function, $\mcf$ \citep{Sheth2004}.
$\mcf$ is a generalization of the two-point correlation function with each galaxy weighted by some mark $m$.
It is defined as 
\begin{equation}
	\mcf = \frac{1}{N_\mathrm{p}(s) \bar{m}^2} \sum_{ij} m_i m_j ~,  
\end{equation}
where the sum is over all pairs with separation $s = s_{ij}$, $N_\mathrm{p}$ is the number of galaxy pairs at $s$, and $\bar{m}$ is the mean of the marks.
Following \cite{WhitePadmanabhan2009}, we choose the marks to be a function of the galaxy number density $\rho_i$ around  galaxy $i$, computed within a sphere of radius $10 \, \hMpc$.
Specifically, we use a mark of $m_i = [\rho_* + \bar{\rho}/(\rho_* + \rho_i)]^n$, where $\bar{\rho}$ is the mean density, following \cite{White2016} and \cite{satpathy_measurement_2019}. 
This mark tends to unity for $\rho \sim \bar{\rho}$, is less than unity for $\rho > \bar{\rho}$ and greater than unity for $\rho < \bar{\rho}$, serving to upweight underdense regions and downweight overdense regions.
The parameters $\rho_*$ and $n$ control the sharpness of the transition.
We test a grid of  $\rho_*$ and $n$ values and choose the values that balance two criteria.
We first select three unique cosmology+HOD models that have a minimal distance between their measured $\wprp$ values.
We then measure $\mcf$ for these catalogs on a grid of varying $\rho_*$ and $n$ values, and see which values maximize the distance between their $\mcf$ values, compared to the variance of the entire test set.
The idea is that we want $\mcf$ to discriminate between models that are indistinguishable with just $\wprp$.
We also want to maximize the variance of the $\mcf$ values overall, so that the predictions can be better distinguished.
These criteria prefer different directions along the $\rho_*$ and $n$ axes, and we choose the values that optimally balance both of them: $n=1$ and $\rho_*=8 \: \bar{\rho}$.

We measure $\mcf$ with the same binning we did $\wprp$, $\cfm$, and $\cfq$, from $s = 0.1$ to $50 \, \hMpc$.
We compute the marks using our modified $k$-d tree code, and use \texttt{corrfunc} \citep{SinhaGarrison2019, Sinha2020} to compute the $\mcf$.

\section{Methods}
\label{sec:methods}

To perform our analysis, we first construct a Gaussian process emulator for each observable, as explained in \S\ref{sec:gp}.
Our inference will require the covariances between the observables and bins; we describe this computation in \S\ref{sec:cov}.
We finally perform the inference using our emulator in combination with Markov Chain Monte Carlo, discussed in \S\ref{sec:inference}.

\subsection{Gaussian process emulation}
\label{sec:gp}

We use a Gaussian process to emulate the function relating the input cosmological, HOD, and assembly bias parameters to the observables.
A Gaussian process is a collection of random variables for which any finite subsample is Gaussian distributed.
It can be described as a multivariate normal distribution generalized to infinite dimensions.
Here, we follow the notation of \cite{RasmussenWilliams2006}; a full discussion of GPs can be found in that text.

Given a training set with $N_\mathrm{train}$ inputs, each with $N_\mathrm{param}$ features $\bm{x}$ and a scalar output $y$, we can construct a design matrix $X$ of shape ($N_\mathrm{param}$, $N_\mathrm{train}$) and a target vector $y$ of length $N_\mathrm{train}$.
We also have a test set with $N_\mathrm{test}$ inputs $\bm{x}_*$ from which we can similarly construct a design matrix $X_*$ and a target vector $y_*$.
We assume that these observations can be described by a function $f$, such that $y = f(\bm{x}) + \epsilon$, where $\epsilon$ is a noise model given by $\epsilon \sim \mathcal{N}(0, \sigma_n^2)$.

The Gaussian process is a function $f(\bm{x})$ relating the input parameters to the output targets.
We take it to have zero mean without loss of generality, and a covariance of $k(\bm{x}, \bm{x'})$, described by a kernel function $k$. 
Extending this to our full design matrices for the training set and including the noise model, the covariance on the targets becomes $\mathrm{cov}(\bm{y}) = K(X,X) + \sigma_n^2\,I$.
We can define the joint distribution of the training target values $\bm{y}$ and the function evaluated at the test inputs $\bm{f}_*$ as
\begin{equation}
    \begin{bmatrix}
    \bm{y}\\
    \bm{f}_*
    \end{bmatrix}
    \sim \mathcal{N}\left( \bm{0}, 
    \begin{bmatrix}
    K(X,X) + \sigma_n^2\,I & K(X,X_*)\\
    K(X_*,X) & K(X_*,X_*)
    \end{bmatrix}
    \right) ~,
\end{equation}
where $K(X,X_*)$ is the covariance matrix of the training and test set inputs, and the other covariances are defined similarly.

Then we can define the predictive function $\bm{f}_*$  as
\begin{equation}
\label{eq:gp_pred}
    \bm{f}_* | X, \bm{y}, X_* \sim \mathcal{N}\left( \bar{\bm{f}}_*, \mathrm{cov}(\bm{f}) \right)
\end{equation}
where the mean $\bar{\bm{f}}_*$ is defined as
\begin{equation}
    \bar{\bm{f}}_* = K(X_*, X) [K(X,X) + \sigma_n^2\,I]\inv \, \bm{y} 
\end{equation}
and the covariance  $\mathrm{cov}(\bm{f}_*)$ as
\begin{multline}
    \mathrm{cov}(\bm{f}) = K(X_*, X_*) \\ 
    - K(X_*, X)[K(X,X) + \sigma_n^2\,I]\inv K(X,X_*) ~.
\end{multline}

Next, we must choose our kernel function, which describes the expected properties of the function we are trying to learn.
We assume the kernel to have only a dependence on the distance between the inputs in parameter space, $r = |\bm{x} - \bm{x'}|$ \new{(i.e., a ``stationary'' kernel)}. 
We test common kernels and combinations, and choose the one that performs the best on our test set:
\begin{equation}
    k(r) = k_\mathrm{exp}(r)\,k_\mathrm{const}(r) + k_\mathrm{M3/2}(r)
\end{equation}
where $k_\mathrm{exp}(r)$ is the exponential squared kernel,
\begin{equation}
    k_\mathrm{exp}(r) = \mathrm{exp}\left( -\frac{r^2}{2l^2} \right) ~,
\end{equation}
$k_\mathrm{const}$ is a constant kernel,
\begin{equation}
    k_\mathrm{const}(r) = c ~,
\end{equation}
and $k_\mathrm{M3/2}$ is a special case of the general Mat\'ern kernel with $\nu=\frac{3}{2}$,
\begin{equation}
    k_\mathrm{M3/2}(r) = \left( 1 + \frac{\sqrt{3}r}{l} \right) \mathrm{exp}  \left( -\frac{\sqrt{3}r}{l} \right) ~,
\end{equation}
where $l$ is a characteristic length scale, and $c$ is a constant.

We train the GP on our set of training catalogs to determine the \new{$2 \, N_\mathrm{param}+1$ kernel parameters (the length scale $l$ for each input parameter for each of the $ k_\mathrm{exp}$ and $k_\mathrm{M3/2}(r)$ kernels, and the constant $c$ for the $k_\mathrm{const}$ kernel)} that result in the maximization of the log marginal likelihood:
\begin{multline}
    \mathrm{log}\,p(\bm{y}|X) = -\frac{1}{2} \bm{y}\T (K + \sigma_n^2\,I)\inv\bm{y} \\
    - \frac{1}{2} \, \mathrm{log}|K + \sigma_n^2\,I| - \frac{n}{2} \, \mathrm{log}2\pi ~.
\end{multline} 
\new{We perform this optimization using the L-BFGS-B solver \citep{fletcher_practical_1987} through \texttt{scipy}.}
We can then use these \new{optimized parameters} to evaluate the kernels in Equation~\eqref{eq:gp_pred}, and use it to predict the target value for our test set inputs.

We train a separate GP model for each bin of each observable.
To perform the Gaussian process computations, we use the \texttt{george} code \citep{Ambikasaran2016}, which is optimized for large data sets.

\subsection{Covariance matrix construction}
\label{sec:cov}

\begin{figure*}
\centering
\subfloat[\label{fig:cov_aemulus}]{\includegraphics[width=0.33\textwidth]{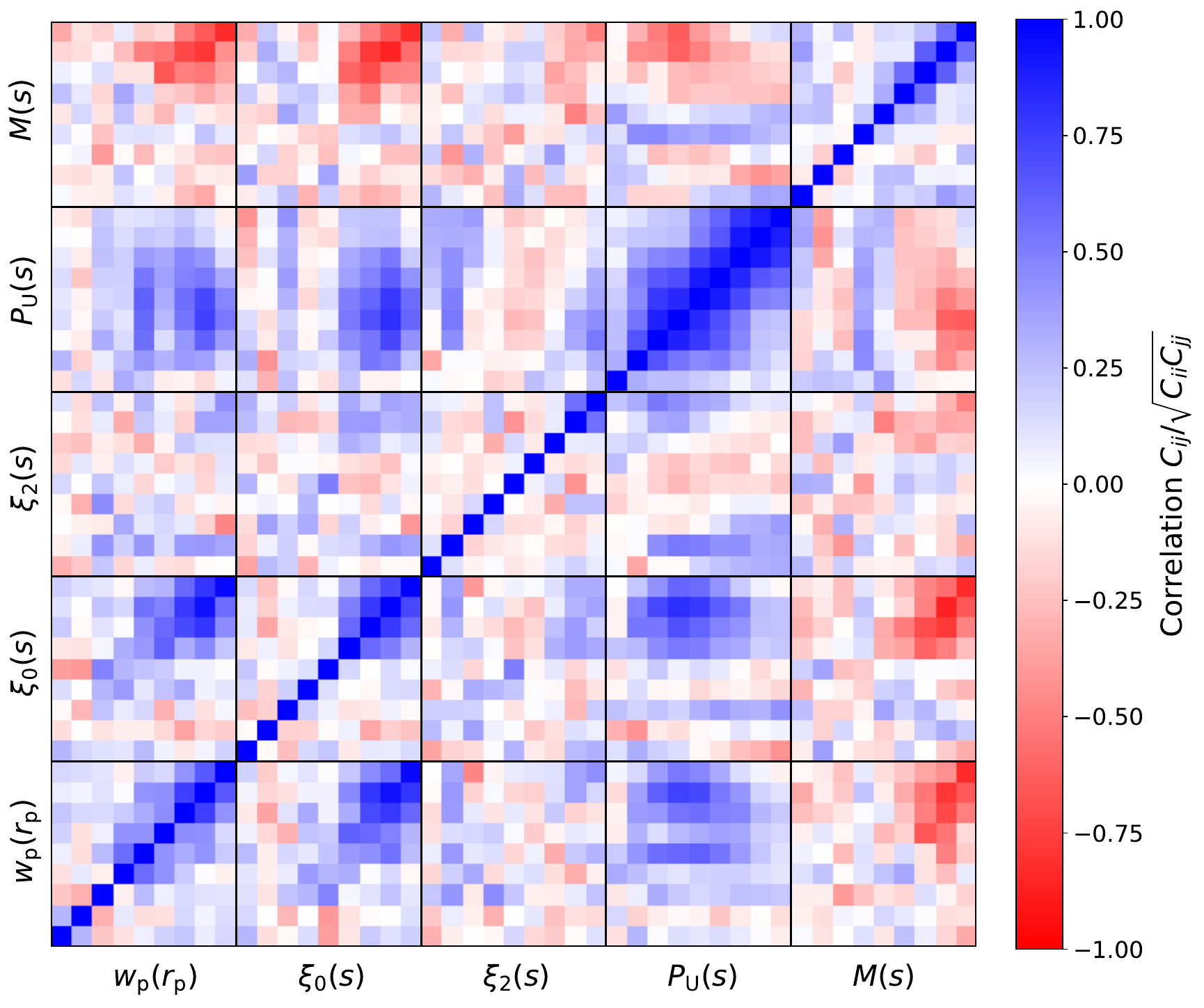}}
\subfloat[\label{fig:cov_emuperf}] {\includegraphics[width=0.33\textwidth]{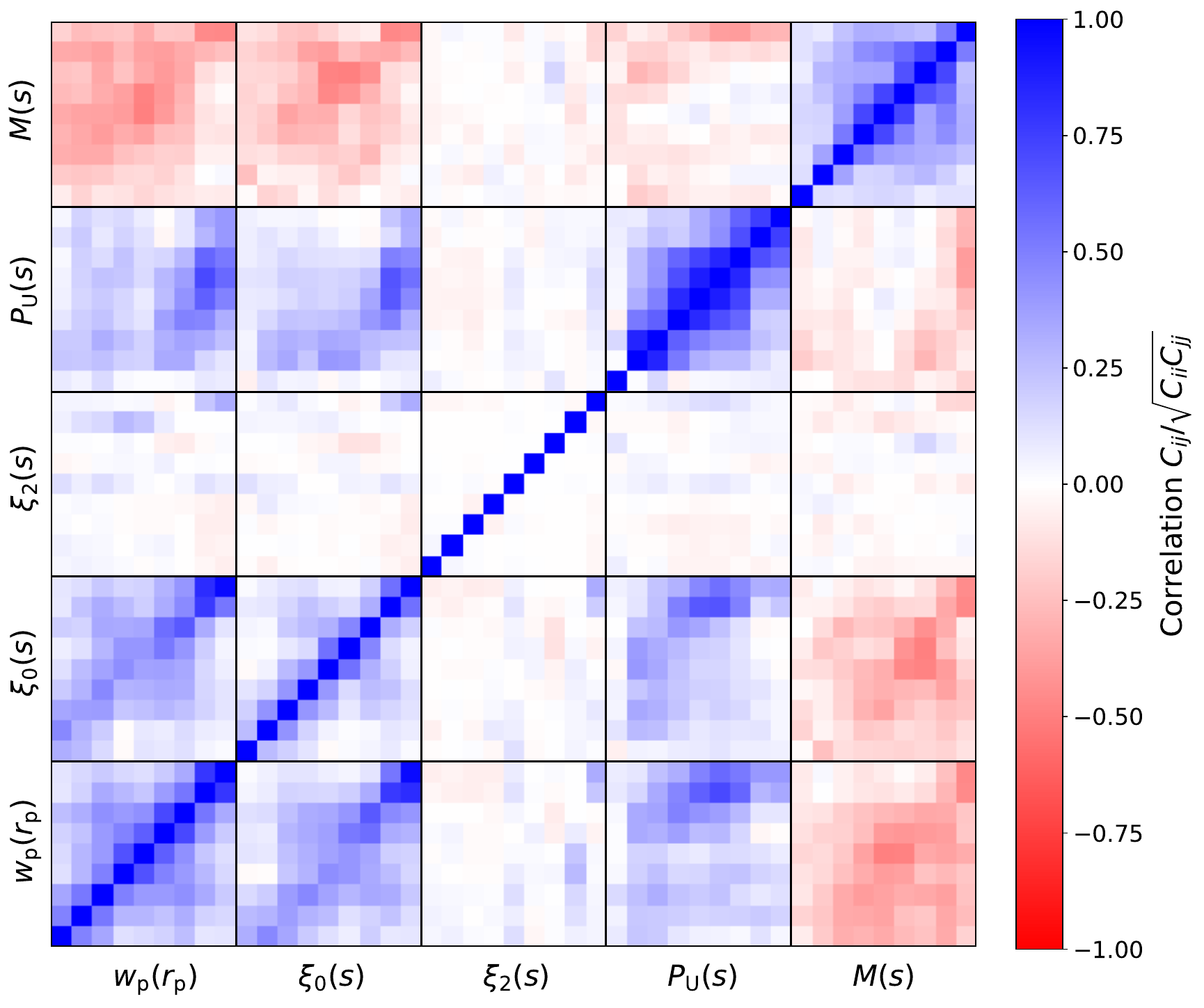}}
\subfloat[\label{fig:cov_smooth_emuperf}]{\includegraphics[width=0.33\textwidth]{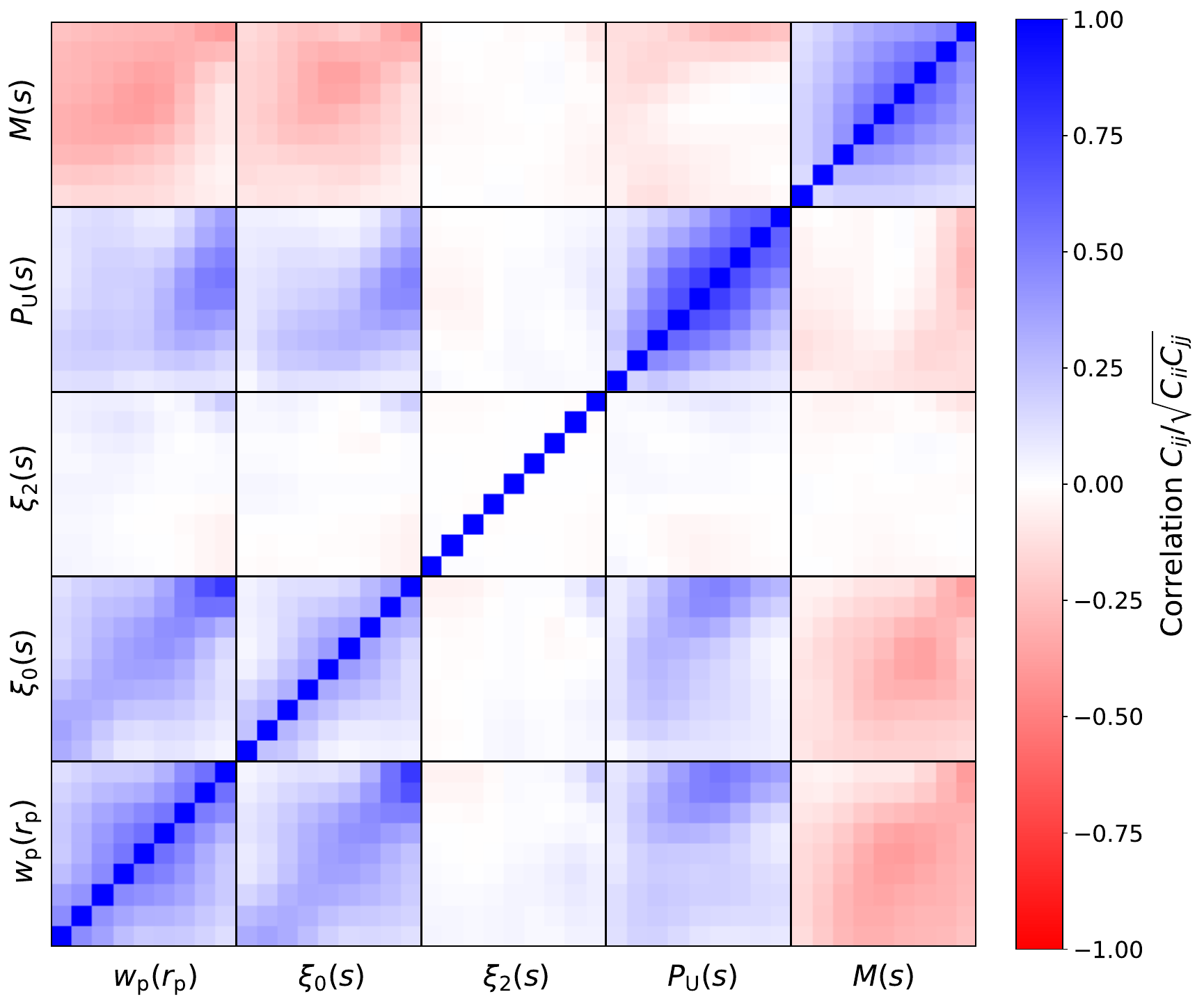}}
\caption{Correlation matrices for visualizing the covariance matrices used in the analysis, for all five observables. The panels show correlation matrices constructed from (a) the \aemulus sample covariance $\cov{aemulus}$, (b) the emulator performance covariance $\cov{perf}$, and (c) the performance covariance with a Gaussian smoothing $\cov{perf,smooth}$. The color bar shows the correlation quantity $C_{ij}/\sqrt{C_{ii}C_{jj}}$, where $C_{ij}$ are elements of the correlation matrix.}
\label{fig:covs}
\end{figure*}

To perform inference using our emulator, we require a covariance matrix describing the correlations between the observables, as well as between the bins of a single observable.
This covariance includes both the uncertainties introduced by the emulator, contained in $\cov{emu}$, and the sample variance of the data on which we are performing parameter recovery, $\cov{data}$.
We combine these into the total covariance $\covtot$ that we will use in our likelihood function (see \S\ref{sec:inference}),
\begin{equation}
    \covtot = \cov{emu} + \cov{data}.
\end{equation}

We define the overall emulator performance covariance $\cov{perf}$ as the combination of both the intrinsic emulator prediction error ($\cov{emu}$) and the covariance of the data on which the emulator is tested, $\cov{test}$, so to obtain $\cov{emu}$ we must subtract off $\cov{test}$:
\begin{equation}
     \cov{emu} = \cov{perf} - \cov{test}.
\end{equation}
We obtain $\cov{perf}$ by computing the covariance of the fractional error between the emulator predictions and the measurements on the data (and then smoothing this matrix to handle noise from our limited number of simulations, as described below).
The performance covariance on our test set with $N_\mathrm{test}=3500$ observations indexed by $n$ is then
\begin{eqnarray}
    \label{eq:frac_pred}
    \cov{perf} &=& \frac{1}{N_\mathrm{test}-1} \sum_n^{N_\mathrm{test}} \bm{f}_n \cdot \bm{f}_n\T ~, \\
    \bm{f}_n &=& \frac{ \bm{y}_{n,\mathrm{pred}} - \bm{y}_{n,\mathrm{test}} }{ \bm{y}_{n,\mathrm{test}} },
\end{eqnarray}
where $\bm{y}$ is a vector of the measured observables (which can be a concatenation of multiple observable vectors).
It is worth noting that we know the expectation value of these fractional errors should be zero, so we assume $\bar{\bm{f}}_{n}=0$ when computing the covariance.
The computed $\cov{perf}$ is visualized in Figure~\ref{fig:cov_emuperf}, for all five observables.

We compute $\cov{test}$ using the \aemulus test set, which has $N_\mathrm{cosmos}=7$ different cosmologies $c$, and $N_\mathrm{box}=5$ boxes (realizations) $b$ for each cosmology. 
These are each populated with $H=100$ HOD models $h$. 
We utilize the fact that we have multiple boxes per cosmology to estimate the sample variance.
We choose a single HOD model in the middle of the parameter space, and for each cosmology populated with this HOD, we compute the mean value of the observable $\bar{\bm{y}}_{c}$ of the $N_\mathrm{box}$ boxes:
\begin{equation}
    \bar{\bm{y}}_{c} = \frac{1}{N_\mathrm{box}} \sum_b^{N_\mathrm{box}} {\bm{y}_{b,c}} .
\end{equation}
We compute the fractional deviation from this mean $\bm{d}_{b,c}$ for each of box of a given cosmology:
\begin{equation}
    \bm{d}_{b,c} = \frac{ {\bm{y}_{b,c} - \bar{\bm{y}}_{c}} } {\bar{\bm{y}}_{c}} .
\end{equation}
We finally compute the covariance of these deviations from the mean:
\begin{equation}
    \cov{aemulus} = \frac{1}{N_\mathrm{box}N_\mathrm{cosmos}-1} \sum_{b}^{N_\mathrm{box}} \sum_{c}^C \bm{d}_{b,c} \cdot \bm{d}_{b,c}\T ~.
\end{equation}
The computed $\cov{aemulus}$ is shown in Figure~\ref{fig:cov_aemulus}.

When we compute Equation~\eqref{eq:frac_pred} used in $\cov{perf}$, the observable values $\bm{y}_{n,\mathrm{test}}$ we use are the mean value of the observable over the $N_\mathrm{box}$ test boxes for each cosmology.
This essentially increases the volume of the test set by a factor of $N_\mathrm{box}$, and uncertainty scales in inverse proportional to volume \citep{KlypinPrada2018}.
Thus, in order to combine $\cov{perf}$ and $\cov{test}$, we need to scale the latter to match the effective volume of the former:
\begin{equation}
    \cov{test} = \frac{1}{N_\mathrm{box}} \, \cov{aemulus} .
\end{equation}

We can now use $\cov{test}$ to construct $\cov{emu}$, and combine it with $\cov{data}$ to obtain the total covariance.
For our tests, we are performing parameter recovery on the \aemulus\ test simulations themselves, so we have $\cov{data} = \cov{test}$, and we get simply $\covtot = \cov{perf}$.
In future applications to real data, we will need to include both $\cov{data}$ and $\cov{test}$ in the covariance matrix construction.

We do use the \aemulus covariance $\cov{test}$ as input to the Gaussian process emulator.
The GP requires an estimation of the uncertainty on the training set.
As the training and test sets are from the same simulation suite, but the test set contains multiple realizations of the same cosmology, we use the test set to estimate the training set uncertainty.
We use the diagonal elements of $\cov{test}$ as the variances $\sigma_n^2$ in Equation~\eqref{eq:gp_pred}.

We perform a smoothing on the total covariance matrix, here $\cov{perf}$, to avoid inference issues due to the initially noisy matrix.
Our procedure follows that of \cite{Lange2022}, and has been shown by \cite{Mandelbaum2013} to give essentially the same results as applying the Hartlap correction to unbias the inverse covariance matrix \citep{Hartlap2007}.
We first compute the correlation matrices, with elements given by $C_{ij}/\sqrt{C_{ii}C_{jj}}$, where $C_{ij}$ are the elements of the covariance matrix.
The diagonal elements of the correlation matrix must be equal to 1, as each element is perfectly correlated with itself, and the surrounding elements are typically much smaller, so we start by replacing the diagonal elements with the mean of its four neighbors.
We then apply a basic Gaussian kernel with width one, to smooth the matrix.
Finally, we replace back the diagonal elements.
The smoothed total covariance matrix, $\cov{perf,smooth}$, is shown in Figure~\ref{fig:cov_smooth_emuperf}.
A comparison between using the smoothed and original covariance matrices for parameter inference is shown in Appendix~\ref{appendix:cov}.

\subsection{Inference with Emulator+MCMC}
\label{sec:inference}

We use Markov Chain Monte Carlo (MCMC) to infer the parameters of the mock catalog given the measured statistics, using the trained Gaussian process emulator to predict the statistic at each set of parameters.
For the MCMC process, we use the package \texttt{dynesty} \citep{Speagle2020}, which implements dynamic nested sampling.
Nested sampling is a method for both obtaining posterior values from a likelihood function and estimating the Bayesian evidence \citep{Skilling2006}; dynamic nested sampling improves upon this by varying the number of live points used in the computation \citep{Higson2019}.
While we do not directly make use of the evidence in this work, dynamic nested sampling is faster and more robust than other standard MCMC approaches.
\new{We use an evidence threshold of \texttt{dlogz=0.1}, and check that our chains are converged with respect to this threshold.
We also perform extensive consistency and convergence tests for other MCMC hyperparameters.}

For the HOD and assembly bias parameters, as well as $\gf$, \new{we use a uniform prior given by the training set parameter range,} with an additional constraint on \new{$\mcut$} to be above $10^{11.5} \: M_\odot$.
For the cosmological parameters, we use a multidimensional Gaussian prior defined by the mean and covariance of the cosmology training set parameter space (see Figure 3 in \citealt{DeRose2018}).
We also try a flat prior and a high-dimensional ellipsoid \new{for the cosmological parameters}, and find no change in the results; we choose to use the multidimensional Gaussian to improve the stability and speed of the MCMC runs.

We use a likelihood $\like$ of
\begin{equation}
    \mathrm{ln} \, \like = -\frac{1}{2} \bigg( \frac{\bm{y}_\mathrm{pred} - \bm{y}_\mathrm{test}}{\bm{y}_\mathrm{test}} \bigg)\T \covtot\inv \bigg( \frac{\bm{y}_\mathrm{pred} - \bm{y}_\mathrm{test}}{\bm{y}_\mathrm{test}} \bigg)
\end{equation}
where $\covtot$ is the covariance matrix described in \S\ref{sec:cov}, and $\bm{y}$ is a vector containing the concatenated observables.
Here, $\bm{y}_\mathrm{test}$ are the statistics measured directly on the test set mock catalog on which we are performing parameter recovery, averaged over the $N_\mathrm{box} = 5$ boxes per cosmology and HOD model, and $\bm{y}_\mathrm{pred}$ are the emulator predictions for the observables at the given point in parameter space.

\section{Results}
\label{sec:results}

In this section, we present the results of our emulation and inference on the \aemulus test suite.
We show the emulator performance (\S\ref{sec:emuperf}), the results of recovery tests on a single test model (\S\ref{sec:recovery_single}) and a larger test sample (\S\ref{sec:recovery_statistical}), and an analysis of the scale dependence of our results (\S\ref{sec:scaledep}).

\subsection{Emulator performance}
\label{sec:emuperf}

\begin{figure*}[htp!]
\centering
\includegraphics[width=0.9\textwidth]{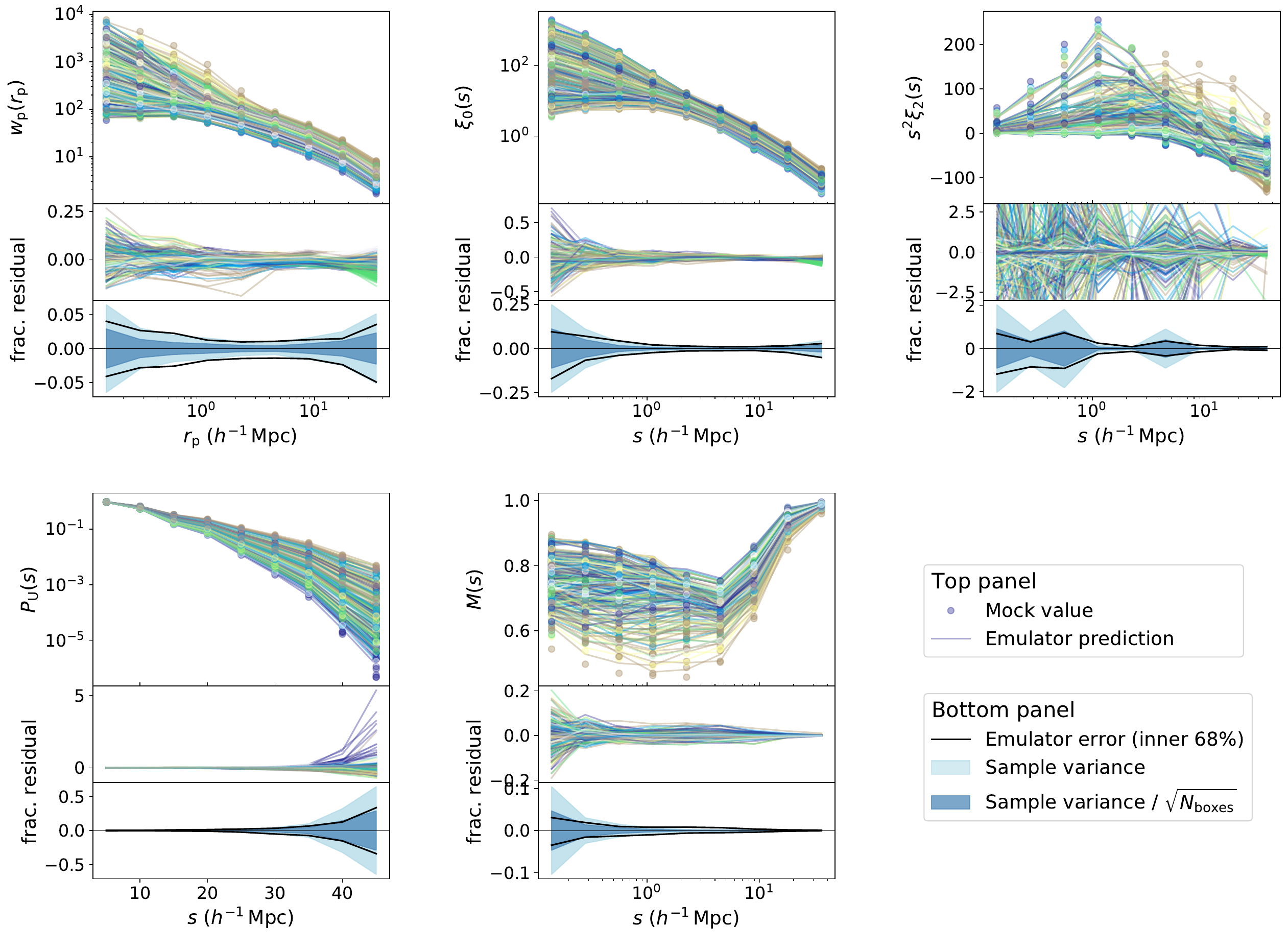}
\caption{The accuracy of our Gaussian process emulator predictions for the projected correlation function $w_{\rm p}(r_{\rm p})$, monopole and quadrupole of the two-point correlation function $\xi_o(s)$ and $\xi_2(s)$, underdensity probability function $P_U(s)$, and marked correlation function $M(s)$. Top panels show the  measured statistics (circles), averaged over $N_\mathrm{box}$ test boxes for each model, and the corresponding emulator predictions (lines) for each cosmology+HOD model. The colors denote different cosmologies. The middle panels show the fractional error of each of the predictions. The bottom panels show the inner 68\% region of the fractional errors (black line), compared to the sample variance of the simulations (light blue). The sample variance scaled by $\sqrt{N_\mathrm{box}}$ adjusts for the effective increase in volume of comparing emulator predictions to the mean of $N_\mathrm{box}$ measurements.}
\label{fig:emu_accuracy}
\end{figure*}

The performance of the emulators is shown in Figure~\ref{fig:emu_accuracy}, for each of the observables for all 700 test models.
For each test cosmology, we compute the statistic for each of the $N_\mathrm{box}=5$ realizations, and take the measured statistic to be the mean of these.
We compute the fractional error between the predicted and measured statistic, and define the error as the symmetrized inner 68\% error.
We compare this error to the sample variance, the square root of the diagonal of $\cov{aemulus}$ for the given observable, as well as this uncertainty scaled by $\sqrt{N_\mathrm{box}}$.
This scaled uncertainty takes into account the increased precision provided by comparing to the mean over multiple boxes; the covariance matrix scales as the inverse volume, as explained in \S\ref{sec:cov}, and averaging over multiple boxes effectively increases the volume, so we obtain this factor of $\sqrt{N_\mathrm{box}}$ (the result is equivalent to taking the square root of the diagonal of $\cov{test}$).

Our emulators achieve very good accuracy across most observables and scales.
For $\wprp$, we obtain \new{$\simo1$--$4\%$, with a minimum at intermediate scales}.
For $\cfm$, we achieve \new{$\simo1$--$2\%$ error on scales between $1$ and $10 \, \hMpc$, and up to $13\%$ for the smallest-scale bin}.
$\cfq$ has the lowest performance, due to high noise levels, \new{with errors ranging from $\simo5\%$ to order unity depending on the scale}.
For $\upf$, we see extremely small errors of \new{$<1\%$ below $20 \, \hMpc$ scales, due to the low variation of the statistic there; up to $35 \, \hMpc$, we achieve $\simo1$--$7\%$ error, with the error increasing even more for the highest-scale bins.}
Finally, for $\mcf$, we achieve $1$--$3\%$ error on scales below 1 $\hMpc$, and $<1\%$ error at larger scales.

At most scales, we see that our emulator error is comparable to \new{the raw sample variance of the \aemulus simulations adjusted for the effective volume. 
The exception is $\wprp$, whose error remains a bit larger than this level at all scales; however, this is not entirely unexpected}, as the emulation performance error includes both the sample variance and the emulator prediction error.

\subsection{Parameter inference recovery tests on a single mock}
\label{sec:recovery_single}

We apply our approach with our GP emulator and MCMC to obtain the posterior distributions of the 18 parameters for a given cosmology+HOD model.
As we have five realizations of each test cosmology, we populate all of these with the same HOD and measure the desired statistics on each of them, and then take the mean of these to obtain the measured statistic. 
These are the values that we compare to the emulator prediction at each step of the MCMC chain.
The \aemulus test volume summed over the 5 boxes is $N_\mathrm{box} \times \, (1.05 \hGpc)^3 = 5.79 \, (\hGpc)^3$.
This is significantly larger than the volume of the highest-redshift shell used in \aemulus V: $1.63 \, (\hGpc)^3$, based on the redshift range $0.48 < z < 0.62$ and the CMASS+LOWZ area of 8447 deg$^2$.
For that analysis, the CMASS data was subsampled to a number density of $2 \times 10^{-4} (\hMpc)^{-3}$, the same as used here, and thus we can make a direct comparison of the volumes. 
The larger volume of the \aemulus test boxes by a factor of a few suggests that these are a meaningful test of the precision we will achieve when we apply the approach to data.

\begin{figure*}
\centering
\subfloat[\label{fig:contour_single_cosmo}]{\includegraphics[width=0.4\textwidth]{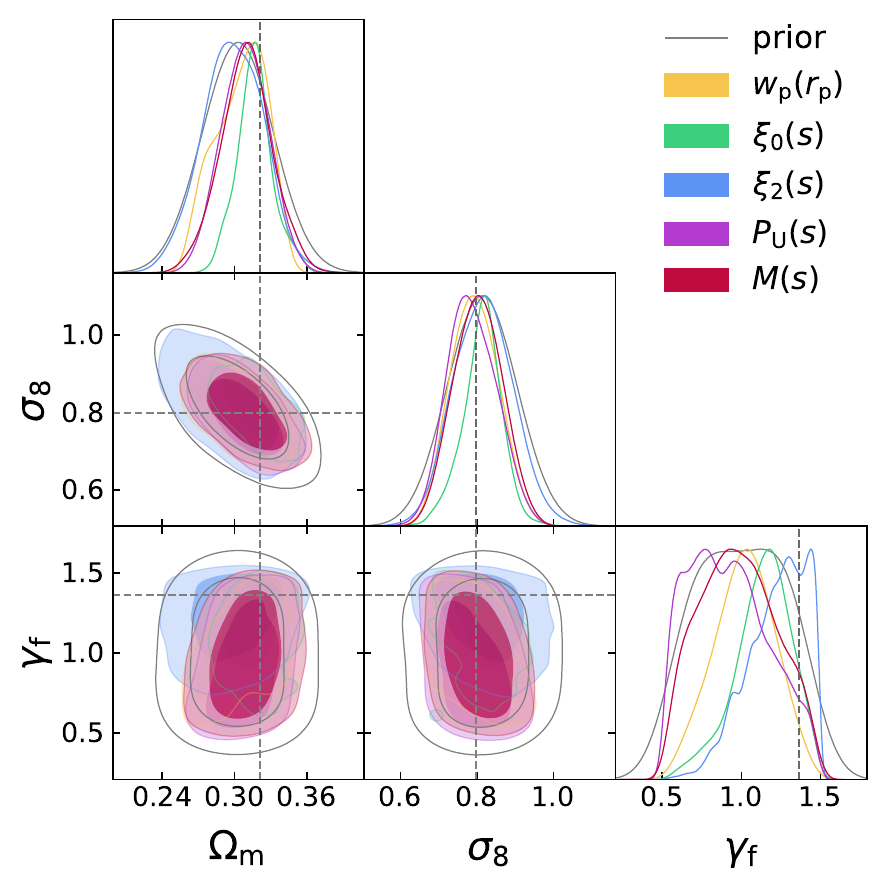}}
\hspace{0.07\textwidth}
\subfloat[\label{fig:contour_single_hodab}] {\includegraphics[width=0.4\textwidth]{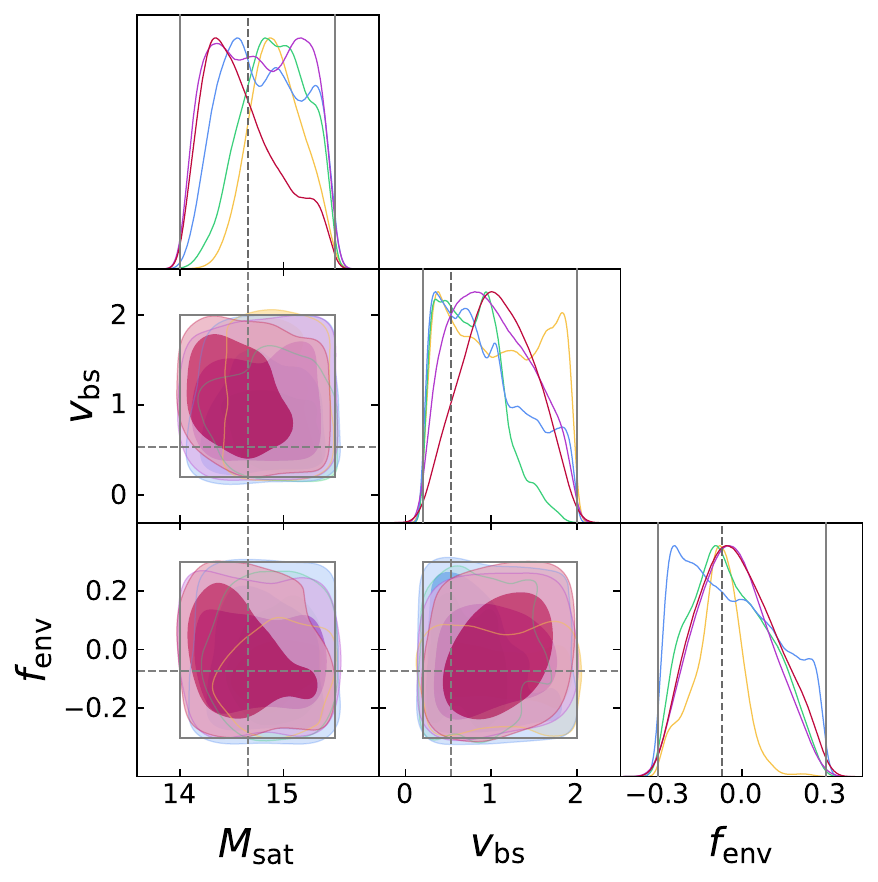}}
\caption{Recovery tests for a single cosmology+HOD model, using a single observable for each MCMC chain. Contours are shown for (a) key cosmological parameters and (b) key HOD and assembly bias parameters.}
\label{fig:contours_single}
\end{figure*}

We start by performing the inference based on each of the five observables alone. 
In Figure~\ref{fig:contours_single}, we show the results on a single cosmology+HOD model; Figure~\ref{fig:contour_single_cosmo} shows key cosmological parameters, and Figure~\ref{fig:contour_single_hodab} shows key HOD and assembly bias parameters.
We have chosen the latter set of parameters because they are particularly degenerate with cosmological parameters.
We see that the different observables have varying effectiveness at constraining the parameters. 
\new{For instance, for this example mock, $\cfm$ provides strong constraints on its own on the cosmological parameters, while the other statistics constrain them more weakly, though $\cfq$ provides surprisingly strong constraining power on $\gamma_f$. 
For the HOD and assembly bias parameters, $\upf$ and $\mcf$ provide constraining power on $\fenv$, though $\wprp$ constrains it even more strongly, and $\cfm$ constrains $\vbs$ well.}
\new{We also note that, because our test set HOD parameter space has the same ranges as our training space, some of the test mocks will have some parameters near the edge of the parameter space.
While this may slightly affect the robustness of the MCMC chains in those regions of parameter space, it applies to a small fraction of the parameters across the mocks and should not affect our overall results.}
\new{Additionally, we note that we have applied optimized smoothing of the posteriors (the default of the \texttt{getdist} plotting software), and this occasionally leads to posterior tails that go slightly beyond the edge of the prior space; we check that the unsmoothed posteriors are entirely within the priors.}

\begin{figure*}
\centering
\subfloat[\label{fig:contour_addin_cosmo}]{\includegraphics[width=0.4\textwidth]{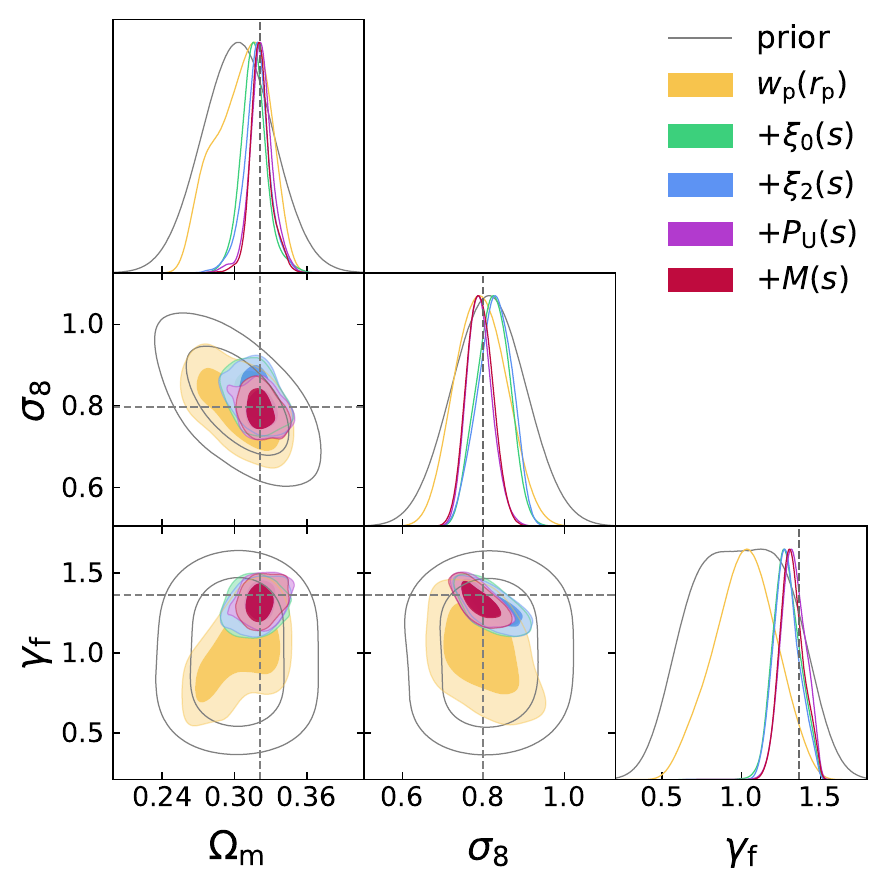}}
\hspace{0.07\textwidth}
\subfloat[\label{fig:contour_addin_hodab}] {\includegraphics[width=0.4\textwidth]{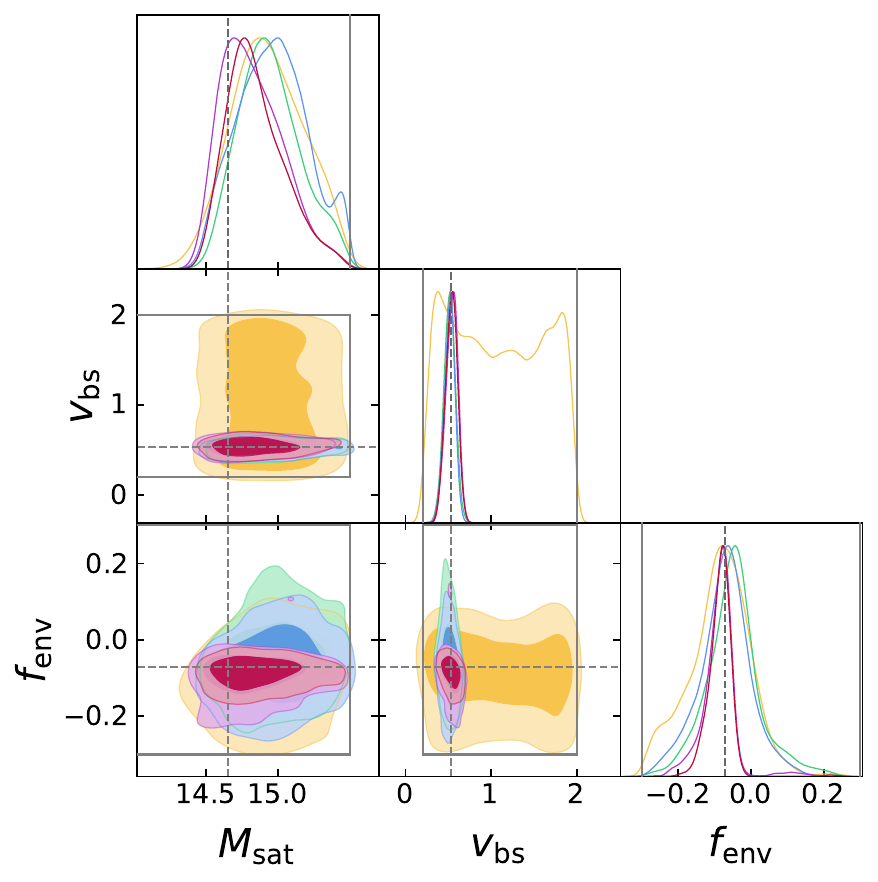}}
\caption{Recovery tests for a single cosmology+HOD model, successively adding in the observables. Contours are shown for (a) key cosmological parameters and (b) key HOD and assembly bias parameters.}
\label{fig:contours_addin}
\end{figure*}

Next, we explore the constraining power of combining the observables when running the MCMC chains.
We start with just $\wprp$, and then one at a time add in $\cfm$, $\cfq$, $\upf$, and $\mcf$.
The results are shown in Figure~\ref{fig:contours_addin} for the same model and parameters as Figure~\ref{fig:contours_single}.
As additional observables are added, we obtain tighter and tighter constraints on the parameters.
In particular, we can compare the constraints with the three standard observables to those when including the two beyond-standard statistics.
\new{For the parameters $\om$, $\sigma_8$, $M_\mathrm{sat}$, and $\fenv$, we see a clear increase in both precision and accuracy when including these new statistics.} 
This indicates promise for the power of the beyond-standard statistics to add additional cosmological information beyond that provided by typical statistics.

\subsection{Statistical results of recovery tests}
\label{sec:recovery_statistical}

\begin{figure*}
\centering
\subfloat[\label{fig:cdf_single}]{\includegraphics[width=0.46\textwidth]{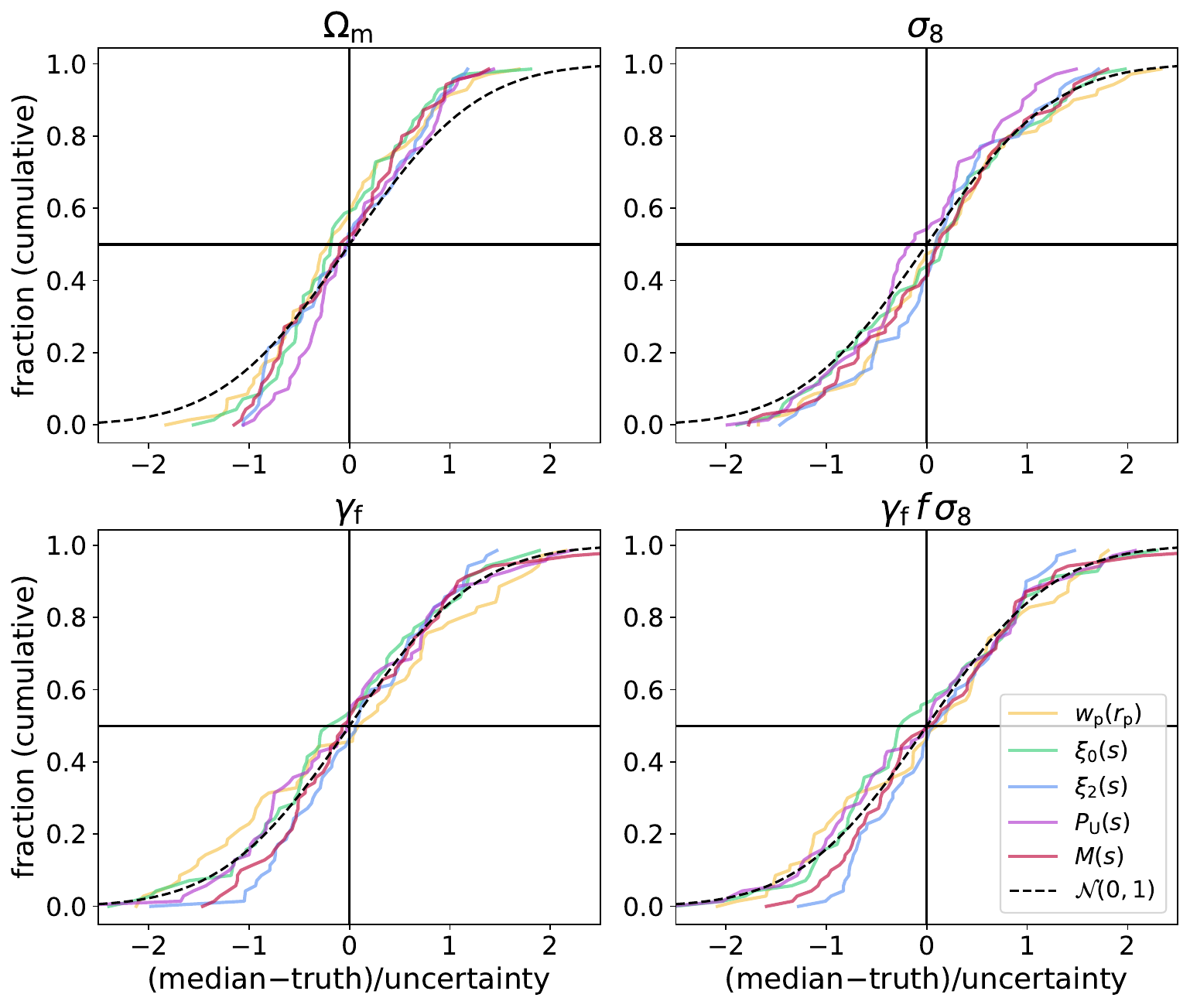}}
\hspace{0.06\textwidth}
\subfloat[\label{fig:cdf_addin_wpmaxscale6}] {\includegraphics[width=0.46\textwidth]{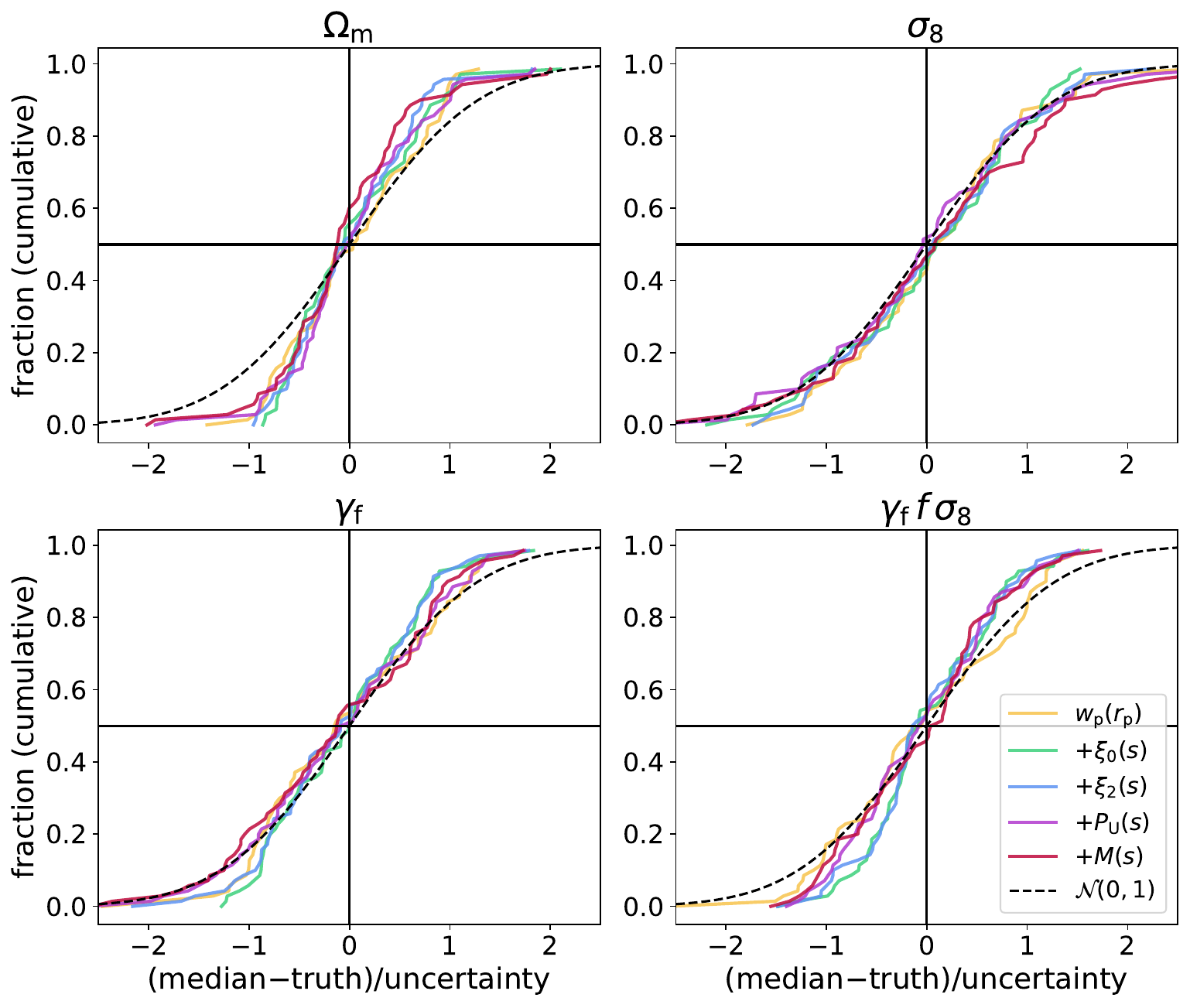}}
\caption{Cumulative distribution functions (CDFs) of the differences between the true parameter value and the median of MCMC chain samples, divided by the uncertainty $\sigma$. Panel (a) shows CDFs for each of the observables on their own, and panel (b) adding in the observables successively; panel (b) excludes the two largest-scale $\wprp$ bins from all combinations, due to a bias discussed in the text. The dashed line shows the CDF of a unit normal distribution for comparison.}
\label{fig:cdf}
\end{figure*}

\begin{figure*}
\centering
\subfloat[\label{fig:recovery_single}]{\includegraphics[width=0.465\textwidth]{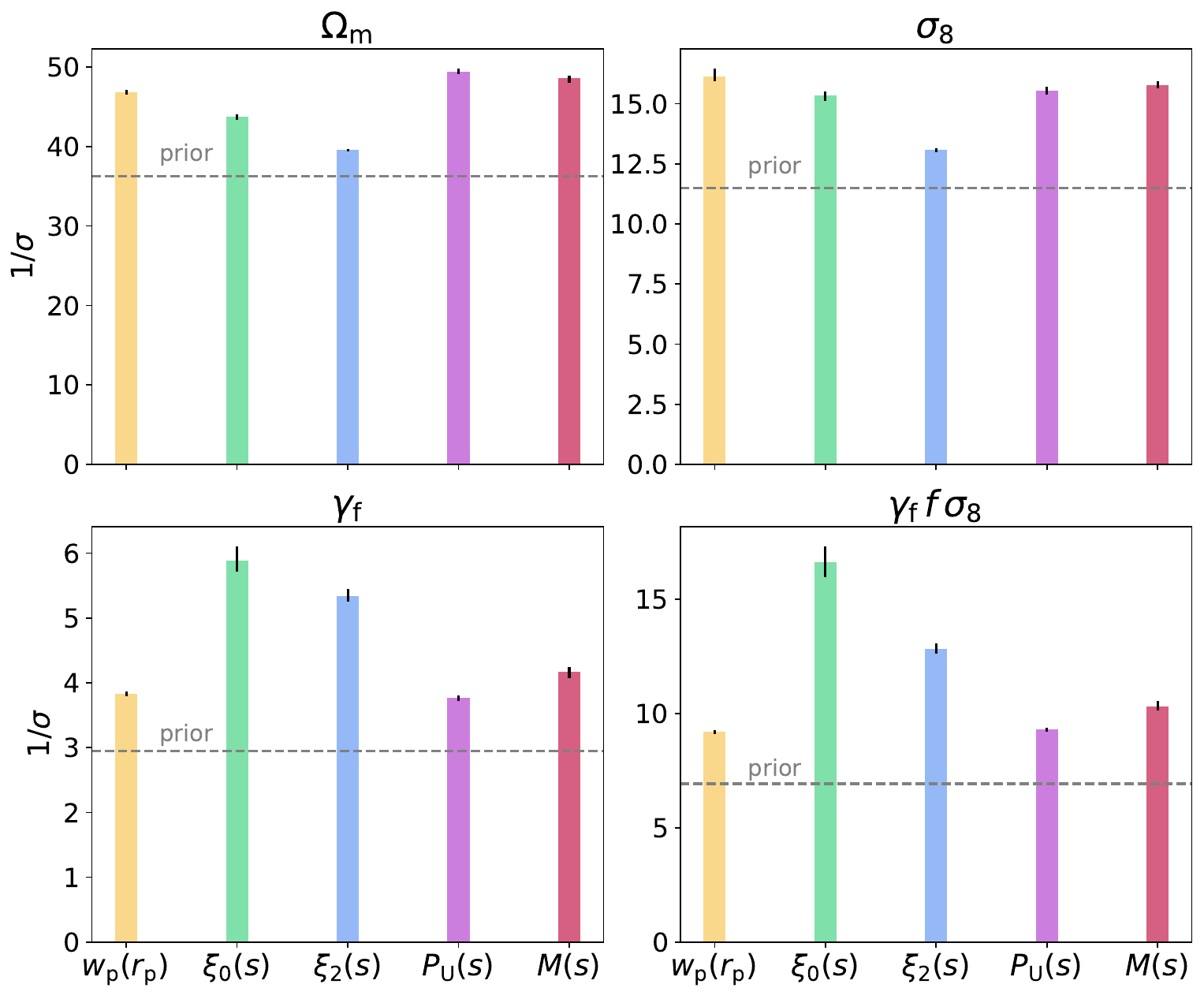}}
\hspace{0.04\textwidth}
\subfloat[\label{fig:recovery_addin}] {\includegraphics[width=0.48\textwidth]{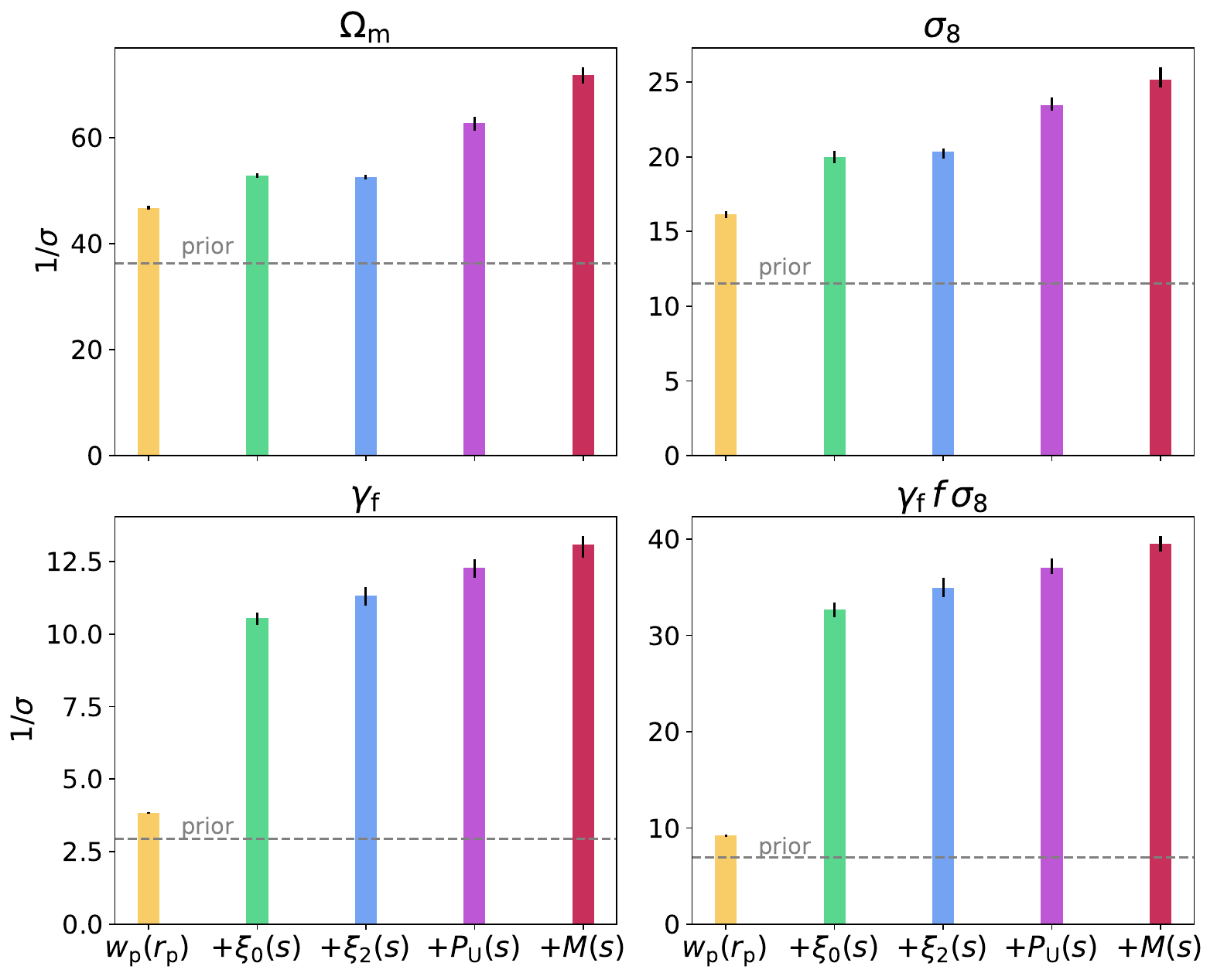}}
\caption{The precision of recovery tests for key parameters, averaged over the 70 test models. The quantity 1/$\sigma$ is the inverse uncertainty on the posterior marginalized over the other parameters, with $\sigma$ defined as the symmetrized inner 68\% region. The precision using only the prior is shown by the gray dashed line. Black bars show the uncertainty on  1/$\sigma$ using bootstrap estimation. Panel (a) shows the precision for tests with single observables, and panel (b) for successively adding in each observable.}
\label{fig:recovery}
\end{figure*}

We perform this MCMC inference for all 70 of our recovery test models (7 cosmologies populated with 10 unique HODs each, averaged over the 5 realizations).
We first assess the accuracy of the recovered parameters by computing the cumulative distribution function (CDF) of the error on the inferred parameter (difference between the median and truth), normalized by the uncertainty, for the 70 recovery test models.
Figure~\ref{fig:cdf_single} shows this CDF for each of the observables used for inference on their own.
We find that, for most of the parameters of interest, the CDF follows a unit normal distribution, which is an indication that the recovery is unbiased.
(We note that the CDF is not an ideal statistic to measure bias, as the function values are dependent on all previous values, but a histogram with only 70 samples is too noisy to make statements about accuracy.)
The exception is $\om$ when using $\wprp$ or $\cfm$; we find that the distribution is biased by $\simo0.5\sigma$ to lower values of $\om$.
\new{There is also a similar slight bias in the $\gfs$ distribution, which we find to be from its dependence on $\om$.}
This bias is small but surprising, as these are both such standard statistics.

To see if the issue could be a result of small number statistics, we run a larger set of recovery tests with $\wprp$ as the sole observable (including all bins), using the full 700 model test suite (each of the seven cosmologies populated with the same 100 HOD models).
We compute the CDF of these 700 results and see that the same bias toward low $\om$ values persists.
With this larger sample, the histogram is less noisy, and the bias is small but clearly visible in the histogram as well.
One possibility is that there are degeneracies with other cosmological or HOD parameters that contribute to $\wprp$ and $\cfm$ favoring lower $\om$ values, but this is difficult to disentangle.

We \new{further} investigate this issue by excluding successively larger scales of $\wprp$ and $\cfm$ from our analysis, as large-scale clustering should be the most affected by $\om$.
We find that removing the two largest-scale bins, above 12.5 $\hMpc$ (with logarithmic averages of 17.7 and 35.4 $\hMpc$) results in an unbiased CDF of recovered $\om$ values.
We check the effect of this bias on the precision of the recovered parameters by rerunning our recovery tests excluding the two largest-scale bins for $\wprp$ and $\cfm$ (but including these bins for the other observables that use them).
We find that, when excluding these scales, the precision we obtain on $\om$ using $\wprp$ decreases by $\simo$13\% and using $\cfm$ by 16\% (averaged over 70 test models); this is similar when using the three standard statistics, and reduces to a precision decrease of only 9\% when including all five statistics.
For the quantity $\gfs$, removing these two bins does significantly decrease the precision by 49\% when using only $\wprp$, but for $\cfm$, this only decreases it by 4\%.
For the three standard statistics, the decrease is 8\%, and for all five, it is only 4\%.
This corresponds to very small changes to our main target result, the relative increased precision when including the beyond-standard statistics compared to the standard statistics, showing that our overall results are robust to this bin exclusion choice. 
The exception is that excluding these two large-scale bins actually increases the relative precision on $\om$, perhaps because the beyond-standard statistics, likely the $\mcf$, are capturing the large-scale information that we are excluding from $\wprp$ and $\cfm$.

For the rest of the results in this paper, except where noted, we exclude the two largest-scale bins for both $\wprp$ and $\cfm$.
We show the CDF when using combinations of successively more observables in Figure~\ref{fig:cdf_addin_wpmaxscale6}.
We find that these distributions are now generally unbiased for $\wprp$, as well as other parameter combinations that include $\wprp$ and $\cfm$, for all of the cosmological parameters;
the CDFs generally follow the unit normal distribution.
Both $\om$ and $\gfs$ show distributions slightly tighter than the normal distribution, indicating that we have overestimated our errors.
This means that our errors may be conservative, but the difference is small and we do not expect this to have significant effects on our results.

We next show our results on the precision of the recovered parameters.
For each parameter, we compute the uncertainty $\sigma$ on the posterior, defined as the symmetrized inner 68\% confidence region, marginalized over the other parameters.
In Figure~\ref{fig:recovery_single}, we show the inverse uncertainty $1/\sigma$ for each of the key cosmological parameters, including the combined quantity $\gfs$, averaged over all 70 test models, when using each of the statistics alone for the inference.
It should be noted that larger bars indicate tighter constraints.
We compare this to the uncertainty obtained when just using the prior.
We see that all of the statistics on their own provide additional constraining power over the prior, for all parameters: \new{$\upf$ provides the most information for $\om$; all the statistics besides $\cfq$ constrain $\sig$ similarly; and $\cfm$ constrains $\gf$ and $\gfs$ the most strongly.}
The amount of information from $\wprp$ is \new{relatively high compared to that} found by other analyses (e.g., \citealt{Lange2022}); we find that this is largely due to our choice to integrate out to only $40\,\hMpc$ along the line of sight, which preserves information in RSDs.
We test integrating out to $80\,\hMpc$ and find much less information content in $\wprp$ alone, though it still contains some.
For the beyond-standard statistics, it is noteworthy that $\upf$ and $\mcf$ do provide information on their own, \new{in particular on $\om$ and $\sig$}. 

We next perform recovery tests adding in each observable one at a time for the full test suite.
We show the results in Figure~\ref{fig:recovery_addin}, again for the mean of 70 test models.
We see that the inverse uncertainty monotonically increases as we add in additional observables.
Our main result is that the constraining power increases significantly between using only the combined standard observables, $\wprp$+$\cfm$+$\cfq$ (blue), and when adding in the beyond-standard statistics as well, $\wprp$+$\cfm$+$\cfq$+$\upf$+$\mcf$ (red).
The change in precision for these two cases tells us the amount of additional information contained in these new statistics: 
The precision increases (defined as the fractional decrease in the uncertainty $\sigma$) by 27\% for $\om$, 19\% for $\sig$, 13\% for $\gf$, and 12\% for the combined growth of structure parameter $\gfs$.
These are significant increases, given the current precision of cosmological measurements.

\subsection{Scale dependence}
\label{sec:scaledep}

\begin{figure*}
\centering
\includegraphics[width=0.9\textwidth]{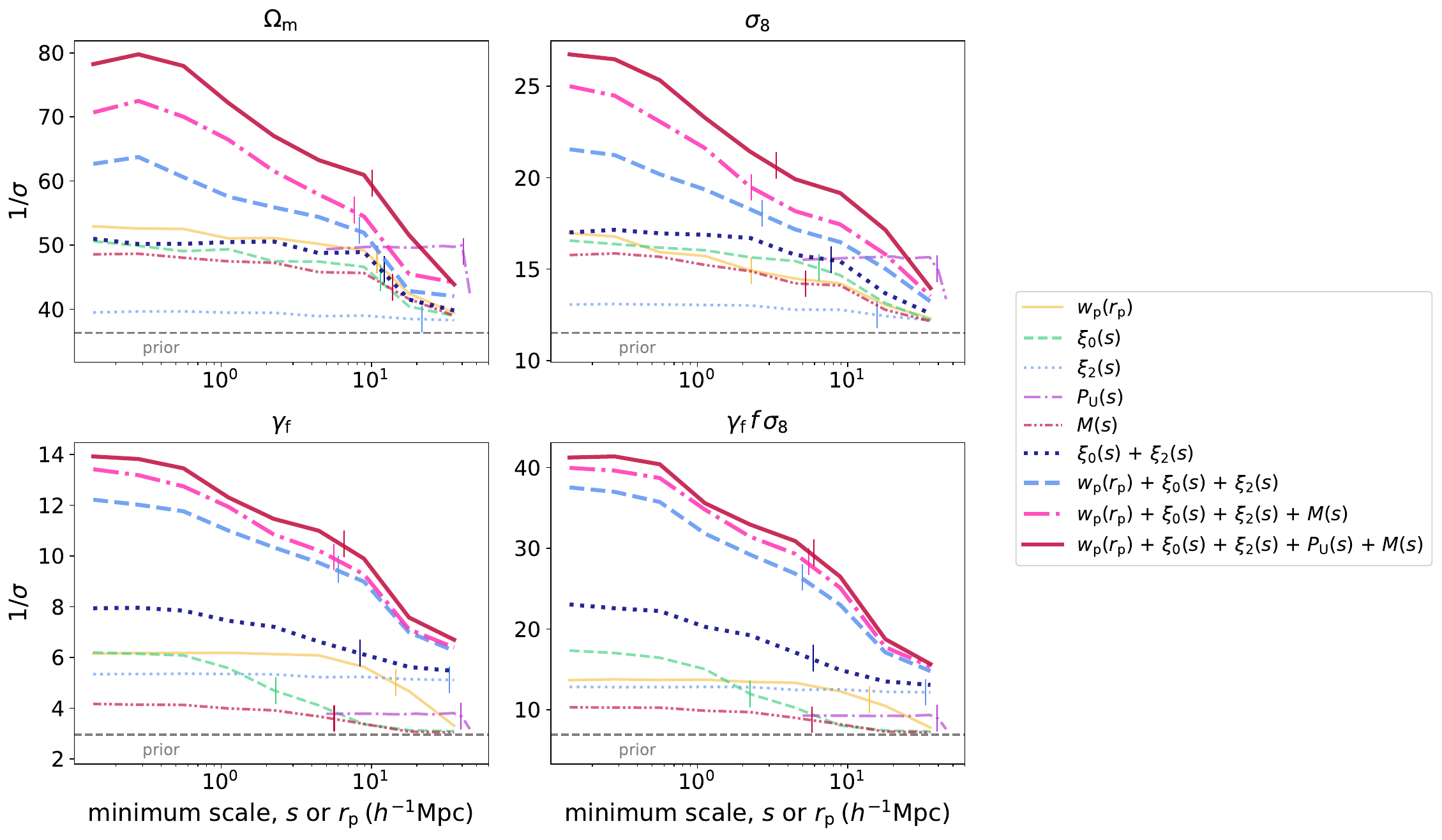}
\caption{The precision of recovery tests as a function of the minimum scale used in the analysis, averaged over the 70 test models. The maximum scale remains fixed at the maximum bin value. The precision is shown for chains using a single observable, as well as for several multiobservable combinations. The vertical bars indicate the scale at which half of the constraining power for that observable is in larger scales and half in smaller scales. We note that $\upf$ is measured on different scales than the other observables, from $5$--$45 \, \hMpc$, so at a minimum scale below $5 \, \hMpc$ it results in an overall shift in precision.}
\label{fig:scale_dependence}
\end{figure*}

We investigate the dependence of our parameter constraints on the scales used in the inference.
To analyze the contribution of small scales, we vary the minimum scale bin used and rerun the MCMC chains, for each parameter individually as well as the five-observable combined constraint.
The results are shown in Figure~\ref{fig:scale_dependence}, averaged over the 70 test models.
We note that the $\upf$ uses a different binning scheme than the other observables, so it is only shown on the scales on which it is computed, $5$--$45 \, \hMpc$, and when it is included in combination with the other observables, it results in an overall shift in precision below $5 \, \hMpc$.
For this reason, we add the $\upf$ and $\mcf$ in the opposite order as the rest of this paper.
We also include using just the combination $\cfm + \cfq$, as many analyses do.
Similarly, we run recovery tests varying the maximum scale.
The $1/\sigma$ lines for the minimum and maximum scale variation will cross each other at a particular scale; this scale is marked by a vertical bar and indicates the scale at which equal information is provided by scales smaller than and larger than this scale.
Thus, a vertical bar far in the small-scale regime means that most of the information comes from small scales (as only the smallest scales are needed on their own to equal the information content in all the larger scales), and conversely, a vertical bar at large scales means that most information comes from large scales. 

As we include smaller scales, the precision increases \new{mostly} monotonically.
Using the vertical bars described above, we find that, for $\gfs$, using the \new{five-observable constraint, scales from $0.1$--$6 \, \hMpc$ provide as much information as the scales $6$--$50 \, \hMpc$}.
\new{We find that the information content continues to increase as we include smaller scales, until a scale of $\simo0.5 \, \hMpc$.}
This is a remarkable finding, given that previous analyses either have not pushed to scales this small or did not find as significant of a contribution from small scales; we discuss this further in \S\ref{sec:discussion}.

To understand this result, we look at the constraints from individual observables for $\gfs$.
\new{For $\cfm$, half of the information comes from scales below $\simo$2.25 $\hMpc$; for $\mcf$, below $\simo$5.75 $\hMpc$; and for $\wprp$, below $\simo15 \hMpc$.
Thus, $\cfm$ is driving the large amount of information on $\gfs$ at small scales, with a contribution from $\mcf$.}
We also look at the constraints on the individual key cosmological parameters $\om$, $\sig$, and $\gf$; \new{for the five-observable constraint, half the information on $\sig$ comes from scales below $\simo$3.3$\hMpc$; on $f$ below $\simo$6.5$\hMpc$; and on $\om$ below $\simo$10$\hMpc$.
Thus, the small-scale constraints on $\gfs$ are driven mainly by the ability of small scales to constrain $\sig$.}
\new{Looking at the commonly used statistic combination $\cfm + \cfq$, we see that the precision nearly flattens out for scales below $\simo 10 \, \hMpc$ for $\om$ and $\sig$.
Including $\wprp$ adds significant constraining power at small scales for all of the parameters; we discuss this further below.}
Finally, adding in $\upf$ and $\mcf$ accesses a significant amount of additional information at smaller scales, in particular for $\om$ and $\sig$.

Notably, the significant additional constraining power from adding in $\wprp$ to $\cfm + \cfq$ differs from the findings of \cite{Lange2022}, who found that it only marginally improved constraints.
Given that the effect of $\wprp$ is \new{somewhat stronger} for $\gf$ and $\gfs$ in our analysis, and \cite{Lange2022} do not include this velocity field rescaling parameter, it seems that the increase in constraining power we find is due to the sensitivity of $\wprp$ to velocity information.
Indeed, we only integrate out to $\pi_\mathrm{max}=40\,\hMpc$, while \cite{Lange2022} uses a value of $\pi_\mathrm{max}=80\,\hMpc$.
We perform a test using this larger value and find that, as expected in this case, $\wprp$ does not add much more constraining power to either $\gf$ or $\gfs$. 
\new{We further investigate this by fixing the $\gamma_f$ parameter in our inference to the true value for each test mock and rerunning the inference, with the goal of isolating the effect of this parameter. 
We find that, when we do this, the difference between the constraints using $\wprp+\cfm+\cfq$ and $\cfm+\cfq$ only is greatly reduced: with free $\gf$, dropping $\wprp$ leads to a 63\% reduction in precision on $\gfs$, while with fixed $\gf$, it is only a 27\% reduction on $\gfs$ (which in this case is determined only by the precision on $f$ and $\sig$).}
Thus, we conclude that our choice of $\pi_\mathrm{max}$ preserves significant velocity information that allows $\wprp$ to constrain the growth of structure parameter through its sensitivity to the halo velocity field rescaling parameter $\gf$.
\new{Though this is a plausible explanation, the information content of $\wprp$ being independent of $\cfm$ and $\cfq$ is still a somewhat surprising result, and the combination of these statistics should be considered carefully in future work.}

\subsection{Recovery tests on external models}
\label{sec:unit}

\new{A key assumption on which our approach relies is that the HOD model is sufficiently flexible to span the space of observed data.
The HOD is just one way of relating halo properties to the galaxy distribution, and incorporates certain (physically and empirically motivated) assumptions about galaxy formation.
This is notable in relation to  perturbation theory approaches, which require a large number of nuisance parameters to model higher-order statistics such as the bispectrum (e.g., \citealt{philcox_cosmology_2022}), while the HOD is a relatively compact parameterization that makes stronger assumptions while showing promise in still being able to model high-order statistics (e.g., \citealt{zhang_constraining_2022}).
To check that our HOD model is sufficiently flexible to model our chosen statistics, we test our approach on a catalog constructed with a different galaxy formation prescription that incorporates different assumptions than the HOD,} namely Subhalo Abundance Matching (SHAM; e.g., \citealt{kravtsov_dark_2004, vale_linking_2004, conroy_modeling_2006}).
This is an important validation step before applying our emulators to real data.
When we adapt our emulators for the full data analysis, we will perform additional tests in this vein to ensure that our framework encompasses the range of expected galaxy formation scenarios.

For this test, we use mock catalogs generated from the \new{UNIT} simulations\footnote{\url{http://www.unitsims.org}} \new{\citep{chuang_unit_2019}} to check that our framework generalizes beyond the \aemulus $N$-body simulations.
\new{The UNIT simulations have a mass resolution of $1.2 \times 10^9 \, h^{-1} M_\odot$ and consist of two pairs of simulations constructed with the fixed-and-paired inverse-phase technique \citep{angulo_cosmological_2016}.
Each simulation has a volume of $(1 \, \hGpc)^3$, leading to an effective volume significantly larger than the \aemulus boxes, so the clustering statistic measurements should be more precise.}
To test that our approach is robust to our use of an HOD model with environment-dependent galaxy assembly bias, we instead populate the \new{UNIT} simulations using the SHAM approach.
SHAM assigns galaxies to subhalos based on a rank-ordered relation between galaxy mass and subhalo mass, with some additional parameters to regulate the scatter, and is able to reproduce galaxy assembly bias to some extent.
We specifically use the SHAM method of \cite{lehmann_concentration_2016} to generate our \new{UNIT} mocks.

For this test, we require a data covariance matrix for the \new{UNIT} mock data, and we use the GLAM Particle-Mesh simulations \citep{KlypinPrada2018} for this purpose.
These have many independent realizations; we use 986 boxes for our covariance estimate.
They are all at the same cosmology; we consider the covariance of the fractional differences from the mean for each statistic.
The GLAM boxes have a volume of $(1 \, \hGpc)^3$, so we rescale the covariance matrix for the \new{effective UNIT} volume.
We use the emulator covariance matrix $ \cov{emu}$ described in \S\ref{sec:cov}, add this to the data covariance, \new{and then perform a Gaussian smoothing on this total covariance matrix} to obtain the final covariance we use in the likelihood function.

\new{We compute the statistics on the UNIT SHAM mocks in the same way as for the \aemulus mocks, first including redshift-space distortions for the UNIT galaxies. 
All of the measured statistics are within 1$\sigma$ of the mean of the training set mocks.
Still, we find that, when we perform a full MCMC over all the parameters with the UNIT statistics, some of the parameter constraints are slightly biased ($\simo$1-2$\sigma$) when adding the two beyond-standard statistics to the data vector.
We note that we also encountered a similar issue when attempting a recovery test using the Uchuu simulations \citep{ishiyama_uchuu_2021} populated with SHAM galaxies.
Below, we will show results based on the UNIT simulation, but the main findings for the SHAM tests hold for Uchuu as well.
This suggests that our HOD parameter space or parameterization may not be sufficiently flexible to encompass the details of the SHAM-distributed galaxies, and that the differences between the HOD and SHAM galaxies manifest in the beyond-standard statistics much more than in the standard statistics.
This is an important issue to investigate further in future work.}

\begin{figure*}
\centering
\includegraphics[width=0.65\textwidth]{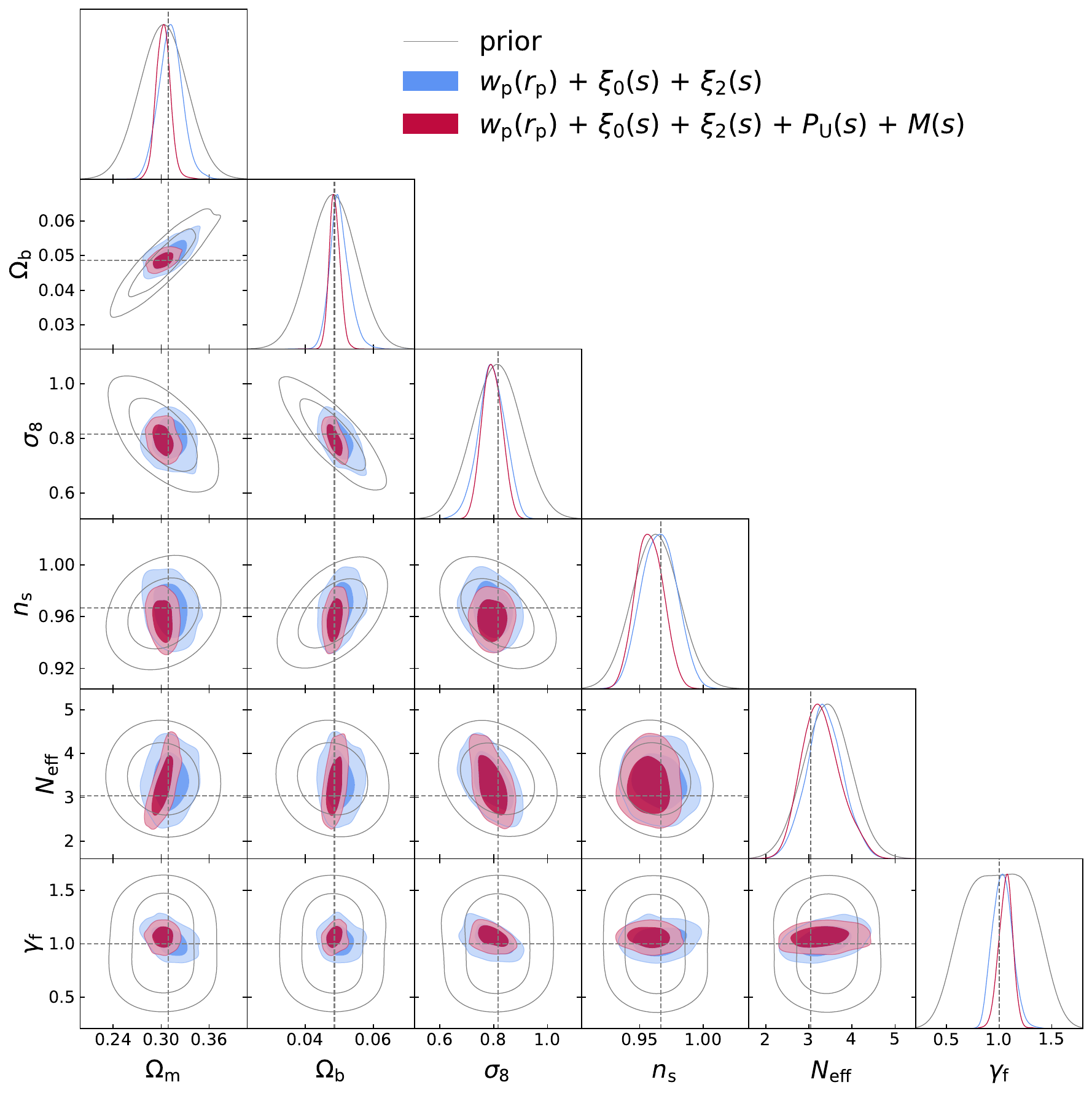}
\caption{Recovery test on the \new{UNIT} mock catalog. Constraints are shown for the cosmological parameters and $\gf$ when using just the standard statistics \new{(blue)}, and when including the $\upf$ and $\mcf$ \new{(red)}; \new{the prior shown for comparison (gray),} and the true parameter values are shown in the dashed gray lines. We keep $w$ and $h$ fixed, as discussed in the text. The parameters are recovered accurately, with the beyond-standard statistics adding increased precision on the parameters of interest.}
\label{fig:unit_recovery}
\end{figure*}

\new{For our SHAM recovery tests, we thus choose to focus our constraints on the growth-related parameters.
We fix two of the cosmological parameters, $w$ and $h$, to their fiducial values (similar to adopting a CMB prior on them), as they are not part of the goals of this analysis (nor other small-scale clustering analyses that use this approach).
We leave all other cosmological parameters and all HOD parameters free in the MCMC exploration.
We also exclude the two largest-scale bins for $\wprp$ and $\cfm$, as these showed a slight bias in the \aemulus recovery test results (see \S\ref{sec:recovery_statistical}).
The results of the tests for the UNIT simulation are shown in Figure~\ref{fig:unit_recovery}, both using just standard statistics and including the beyond-standard statistics.
We find that, in both cases, we can accurately recover the \new{UNIT cosmological parameters as well as $\gf$}.
The inclusion of the beyond-standard statistics results in an increase in precision of 40\% on $\Omega_m$, 25\% on $\sig$, 30\% on $\gf$, and 17\% on $\gfs$; this is similar to (in fact, even better than) our findings with \aemulus recovery tests.
We do note that, for some of the parameters, the results become slightly more biased when adding in the beyond-standard statistics, but all parameters are still recovered to within $1\sigma$.
This may be related to the aforementioned issue with differences between the SHAM and HOD galaxies that are captured by those statistics.
We also test fixing other combinations of cosmological parameters, and find that fixing $w$ and $N_\mathrm{eff}$ (with $h$ and the others free) or fixing $w$ and $\Omega_\mathrm{b}$ produces results similar to the case shown here.
Fixing more than two of these parameters does not change the results, so we chose the two-fixed-parameter test as our fiducial case.
While this SHAM test does reveal caveats to our approach, the results are still promising for the application of this framework to real data sets.}

\section{Discussion and Conclusions}
\label{sec:discussion}

We have constructed Gaussian process emulators for galaxy clustering statistics using the \aemulus simulation suite, including the nonstandard statistics the underdensity probability function $\upf$ and the marked correlation function $\mcf$, which we expect to contain additional information relevant to constraining cosmological parameters of interest.
We achieve typical prediction errors of $\simo2\%$ with our emulator, depending on the scale and statistic.
Using held-out test simulations, we perform recovery tests to determine how well we can constrain the input parameters.
We find that including the beyond-standard statistics significantly increases the precision on the recovered parameters, by 19\% on 
$\sig$, 27\% on $\om$, and 12\% on $\gfs$.
We confirm that our framework is robust to different simulations and galaxy bias models by testing it on mock catalogs constructed from the \new{UNIT} simulations and the SHAM method, on which we achieve unbiased constraints \new{and a similar improvement in precision when including the beyond-standard statistics}.  

To follow this proof-of-concept work, we will apply these emulators to measure the growth of structure in a current galaxy sample (BOSS or DESI). 
We expect that our combination of beyond-standard statistics with small-scale emulation will improve constraints; for instance, \cite{satpathy_measurement_2019} used the marked correlation function to analyze the BOSS data and found that their results were limited by modeling RSD effects on small scales.
This analysis will require a careful treatment of many issues and subtleties in real data. 
We will have to handle redshift evolution, by working in redshift slices with emulators trained at the proper redshift.
We will require a sample constant in number density, both to match our emulators and because void- and density-based statistics are particularly sensitive to variations in number density.
One of the main issues when applying to BOSS data will be fiber collisions, which lead to galaxies without measured redshifts, producing a nontrivial impact on clustering measurements especially at small scales (e.g., \citealt{Zehavi2002}).
Additionally, we will have to handle survey geometry effects including edges and bad fields.
The underdensity probability function and the local density-based marks used for the marked correlation will both be especially sensitive to these issues; we will apply fiber collision weights to the statistics and volume corrections to the spheres used for the density computations, and perform robust tests to ensure that we can recover unbiased parameters.

The application of this work to the BOSS sample will extend the project of \aemulus V \citep{zhai_aemulus_2023}.
The \aemulus V analysis used $\wprp$, $\cfm$, and $\cfq$, the standard statistics discussed in this paper, and obtained tight constraints on the growth of structure parameter $\fsig$ in three redshift bins.
The analysis obtained a low value of $\fsig$ compared to Planck constraints based on a $\Lambda$CDM+GR model, adding to a recent wave of similarly low results based on small-scale clustering \citep{Chapman2021, Lange2022, Yuan2022}.
These studies are also based on standard clustering statistics; bringing in additional statistics and thus additional constraining power will allow for clearer tests of internal consistency between these analyses, as well as testing the demonstrated tension with Planck results.

There are multiple effects that could be contributing to this $\fsig$ tension.
One is additional baryonic effects that influence galaxy formation and are unmodeled in the HOD, introducing errors; while these are unlikely to be relevant at current precision, in future surveys they may become important.
Future work will incorporate additional flexibility in the galaxy bias and assembly bias models to test this hypothesis, and this will in turn require increased constraining power from the data.
The complementary information provided by nonstandard statistics, as shown in this work, will be important in offsetting this flexibility to obtain high-precision constraints on $\fsig$ and help confirm or rule out this explanation for the $\fsig$ tension.
Another potentially relevant effect is that of massive neutrinos, which suppress the growth of structure in a scale-dependent way.
The next generation of the \aemulus simulations \new{\aemulus $\nu$, \citealt{derose_aemulus_2023}} will incorporate massive neutrinos, and the emulation of nonstandard statistics will also be important in obtaining precise small-scale constraints from this updated model.

\new{In this work, we have included a detailed handling of the covariances between the observables, incorporating both the data and emulator covariances in our inference.
To estimate the relative contribution of these sources of uncertainty in our target analysis, we perform a volume-based scaling of the data covariance of the \aemulus test boxes ($\cov{aemulus}$) to one of our target samples, the CMASS high-$z$ bin.
We find that this data covariance is of similar order to the emulator covariance, and the dominating source of uncertainty depends on the observable and scale; in either case, they are never more than a factor of $\simo2$ different.
While this indicates that we are still theory-limited in some regimes, this is reasonable given the newness of the emulator approach.
A comparison between the emulator and data covariances for the standard statistics is also shown in Figure 15 of \aemulus V \citep{zhai_aemulus_2023}, which similarly finds that the errors are comparable.
Future iterations of this type of analysis will be able to reduce the theory uncertainty through a combination of more training simulations (as in \aemulus $\nu$), larger simulation volumes, and improved emulation techniques.
These improvements will become increasingly important as the data uncertainty also gets reduced with future observations.}

The effects of galaxy assembly bias are not yet a concern, given the current precision of our surveys, as shown in the \cite{zhai_aemulus_2023} BOSS RSD analysis, but as both our data and constraining power of methods improve, this will become a key source of uncertainty.
Previous works have found a small but significant dependence on halo environment (e.g., \citealt{Zehavi2018, Yuan2021}).
The density-sensitive statistics we investigate here---namely the $\mcf$ with marks given by the galaxy number density on $10 \, \hMpc$ scales, and the $\upf$, which measures underdense regions across a large range of scales---target this environmental bias.
We have shown that \new{these statistics are well-positioned to improve constraints on cosmological parameters by breaking degeneracies between cosmological and environmental assembly bias effects.}
Other sources of assembly bias, such as halo formation time, concentration, and spin, could be analyzed with marked correlation functions based on these properties or other similarly targeted statistics; these can be readily incorporated into our emulation framework.

More broadly, this work confirms that additional, \new{beyond-standard} clustering statistics\new{, namely the $\upf$ and $\mcf$,} can increase the constraining power in existing data, with little added cost.
This approach could be extended to include other statistics that depend on the goals of the analysis.
These could include the three-point function (e.g., \citealt{TakadaJain2003, McBride2011}), the $k$NN-CDF \citep{Banerjee2021a}, and galaxy group statistics such as the group multiplicity function (e.g., \citealt{BerlindWeinberg2002}) and the group velocity dispersion.
We will explore some of these in future work.
\new{It is important to note that these statistics may be more sensitive to the choice of galaxy bias model than standard statistics, as we found in our initial tests on SHAM galaxies (\S\ref{sec:unit}).
This should be carefully checked when incorporating new statistics; in our case, we do find that including the $\upf$ and $\mcf$ result in slightly biased parameter constraints on the SHAM galaxies when all cosmological parameters are left free.
This may point to the need for an even more flexible HOD parameterization, an investigation we leave for future work.}

One of the primary goals of the \aemulus project is to extract information from small-scale clustering, which is difficult to model theoretically and expensive to simulate fully.
Here, we have shown that there is significant information at small scales for nearly all of the statistics we analyze.
For the constraint on $\gfs$, we find that scales from \new{$0.1$--$6 \, \hMpc$ contribute half of the information content, and that there is additional information all the way down to $0.5 \, \hMpc$.}
This confirms a similar result by \cite{Zhai2019}, which uses $\wprp$, $\cfm$, and $\cfq$, and includes the halo velocity field scaling parameter $\gf$. 
Some recent analyses have not found as much additional information at these small scales.
\cite{Lange2022} conclude that, for their low-redshift sample, which is closer in number density to the one analyzed here, scales between $1$ and $2 \hMpc$ increase the constraining power on $\fsig$ by a small amount, and scales below $\simo1 \hMpc$ not at all.
As discussed in \S\ref{sec:scaledep}, they do not incorporate a $\gf$ parameter to scale the velocity field, and they do not use $\wprp$ as we do.
This model flexibility, which the work of \cite{Zhai2019} also includes, combined with a statistic sensitive to velocity information, may allow us to extract additional information from small scales.
The analysis by \cite{Lange2022} does include an assembly bias model using the decorated HOD framework \citep{Hearin2016}, but this is not as flexible as our three-parameter environmental assembly bias model.
Our increased flexibility on this front may also contribute to the discrepancy, though future work should revisit these hypotheses.

Finally, in this work we built emulators at fixed redshift and scale.
To apply to different data sets, we will require predictions at various redshifts, for which suites of emulators can be constructed and trained at the needed redshifts; an extension of this work could construct emulators that are able to make predictions as a continuous function of redshift.
In a similar vein, here we emulated the clustering statistics at fixed scale, with a different model trained for each bin. 
In future work, we could train the model on all bins simultaneously to include the full covariance properties; even better, we could include scale as an input parameter and make predictions at any scale.

\section*{Acknowledgements}

K.S.F. thanks Sean McLaughlin, Johannes Lange, Sihan Yuan, David W. Hogg, and the Astronomical Data group at the Center for Computational Astrophysics for helpful discussions.
We are grateful to the anonymous referee, whose feedback has greatly improved this manuscript.
This work received support from the U.S. Department of Energy under contract number DE-AC02-76SF00515 and from the Kavli Institute for Particle Astrophysics and Cosmology.
K.S.F. is supported by the NASA FINESST program under award number 80NSSC20K1545.
J.L.T. acknowledges support of NSF grant AST-2009291.
J.D.~is supported by the Lawrence Berkeley National Laboratory Chamberlain Fellowship.
This research made use of computational resources at SLAC National Accelerator Laboratory and the NYU Department of Physics; the authors thank the SLAC computational team and Mulin Ding at NYU for computational support.
This research used resources of the National Energy Research scientific Computing Center, a DOE Office of Science User Facility supported by the Office of Science of the U.S. Department of Energy under Contract No. DE-AC0205CH11231.

We thank Instituto de Astrofisica de Andalucia (IAA-CSIC), Centro de Supercomputacion de Galicia (CESGA), and the Spanish academic and research network (RedIRIS) in Spain for hosting Uchuu DR1, DR2, and DR3 in the Skies \& Universes site for cosmological simulations. The Uchuu simulations were carried out on Aterui II supercomputer at Center for Computational Astrophysics, CfCA, of National Astronomical Observatory of Japan, and the K computer at the RIKEN Advanced Institute for Computational Science. The Uchuu Data Releases efforts have made use of the skun@IAA\_RedIRIS and skun6@IAA computer facilities managed by the IAA-CSIC in Spain (MICINN EU-Feder grant EQC2018-004366-P).

The UNIT simulations were run at the MareNostrum Supercomputer hosted by the Barcelona Supercomputing Center (BSC), Spain, thanks to the PRACE project
grant No 2016163937.

\section*{}
 
\textit{Software:} numpy \citep{VanDerWalt2011}, IPython \citep{Perez2007}, scipy \citep{Virtanen2020}, matplotlib \citep{Hunter2007}, corrfunc \citep{SinhaGarrison2019, Sinha2020}, halotools \citep{Hearin2017}, dynesty \citep{Speagle2020}, george \citep{Ambikasaran2016}, getdist \citep{Lewis2019}.
Our emulator and related data products are publicly accessible in living form at \url{https://github.com/kstoreyf/aemulator} and in archived form at \url{https://zenodo.org/badge/latestdoi/377599274} \citep{storey-fisher_k_githubcomkstoreyfclust_2023}.
The code used to compute the clustering statistics is available in living form at \url{https://github.com/kstoreyf/clust} and in archived form at \url{https://zenodo.org/badge/latestdoi/171347348} \citep{storey-fisher_k_githubcomkstoreyfaemulator_2023}.

\appendix

\section{Covariance matrix comparison}
\label{appendix:cov}

\begin{figure*}[t]
\centering
\subfloat[\label{fig:contour_cov_goodmodel}]{\includegraphics[width=0.47\textwidth]{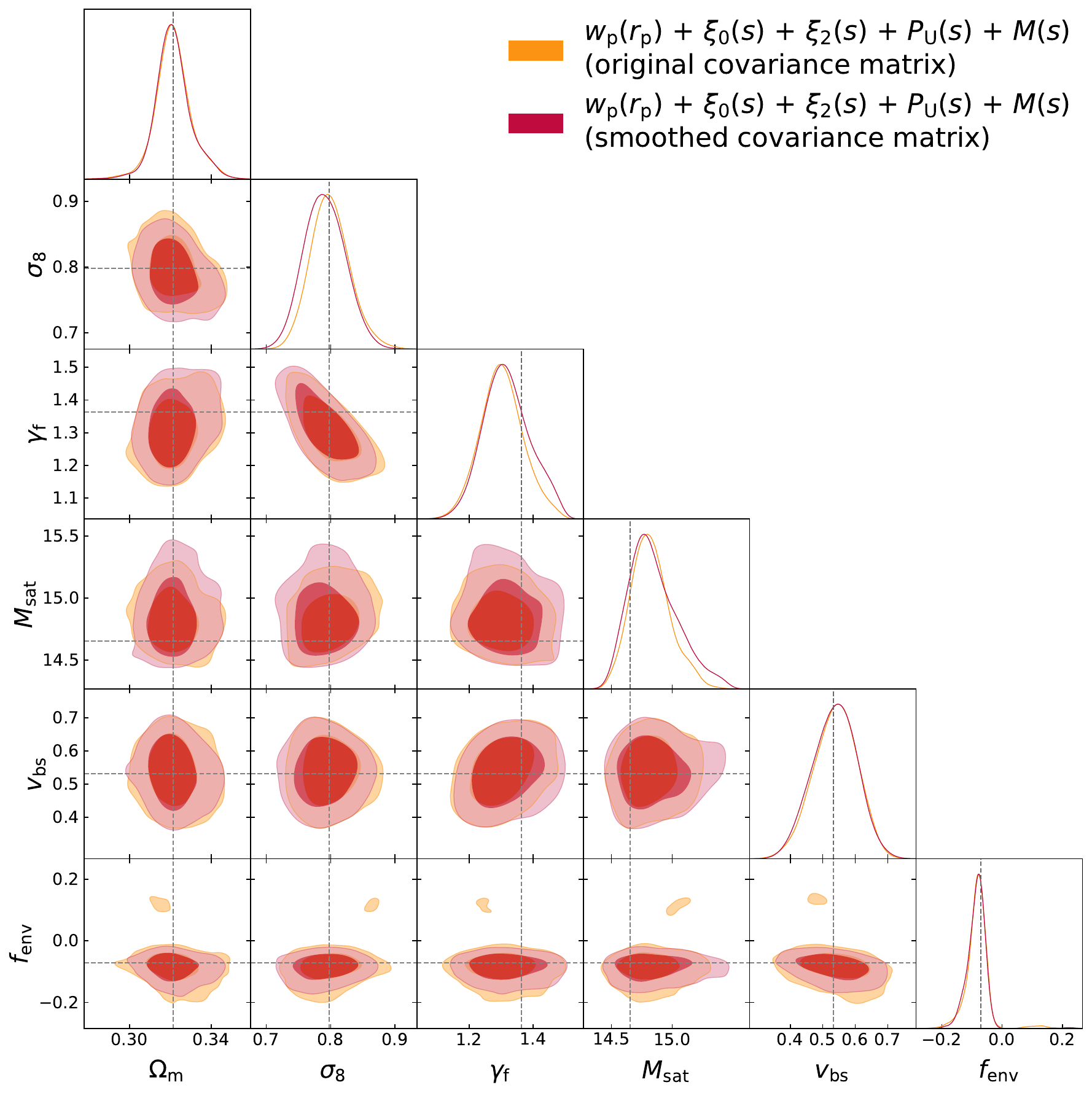}}
\hspace{1em}
\subfloat[\label{fig:contour_cov_poormodel}]{\includegraphics[width=0.47\textwidth]{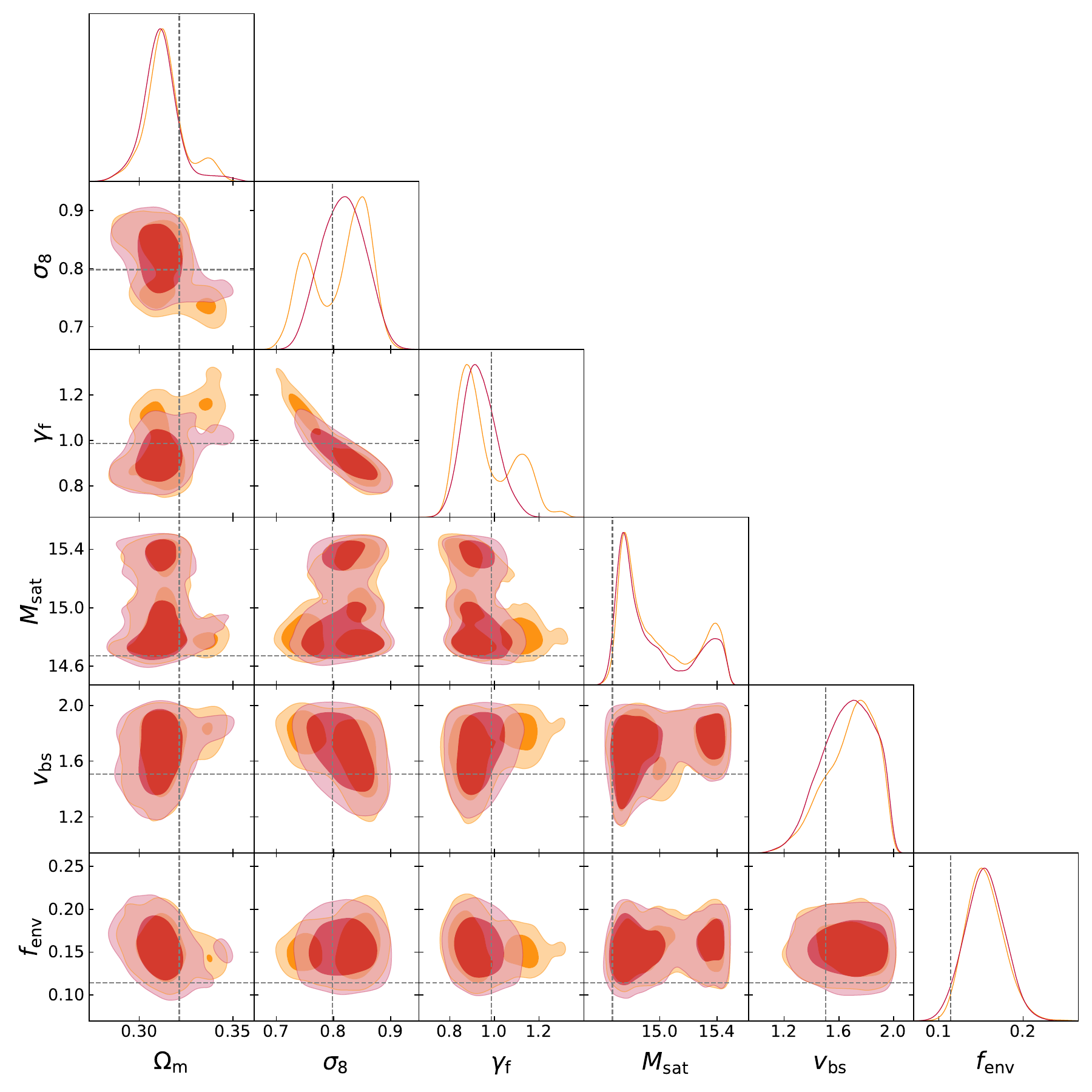}}
\caption{A comparison of the effect of the covariance matrix on recovered parameters. Panels (a) and (b) show recovery tests of key parameters for two different cosmology+HOD models, using all five observables, with the original covariance matrix compared to the covariance matrix with a Gaussian smoothing.}
\label{fig:contour_cov}
\end{figure*}

We compare the posteriors of recovery tests when using the original noisy covariance matrix compared with the Gaussian-smoothed covariance matrix, as described in \S\ref{sec:cov}.
The results are shown in Figure~\ref{fig:contour_cov} for two different cosmology+HOD models for a mix of key cosmological and HOD parameters.
We find that, for the generally well-behaved model (Figure~\ref{fig:contour_cov_goodmodel}), the posteriors are similar between the two covariance matrices, with the smoothed matrix resulting in slightly more accurate parameter estimates.
For the less well-behaved model (Figure~\ref{fig:contour_cov_poormodel}), the posteriors are quite noisy with the original covariance matrix.
Using the smoothed version cleans up some of the spurious modes in the posteriors, suggesting that the smoothing does help in avoiding issues related to noise in the covariance matrix. 
However, some of the modes persist even when using the smoothed matrix, indicating that perhaps we are still not properly sampling our parameter space, or that some of these regions of parameter space may be actual good fits to the observables and indicate true degeneracies in the parameters.

\section{Recovery test results for HOD \& assembly bias parameters}

\begin{figure*}[t]
\centering
\includegraphics[width=0.6\textwidth]{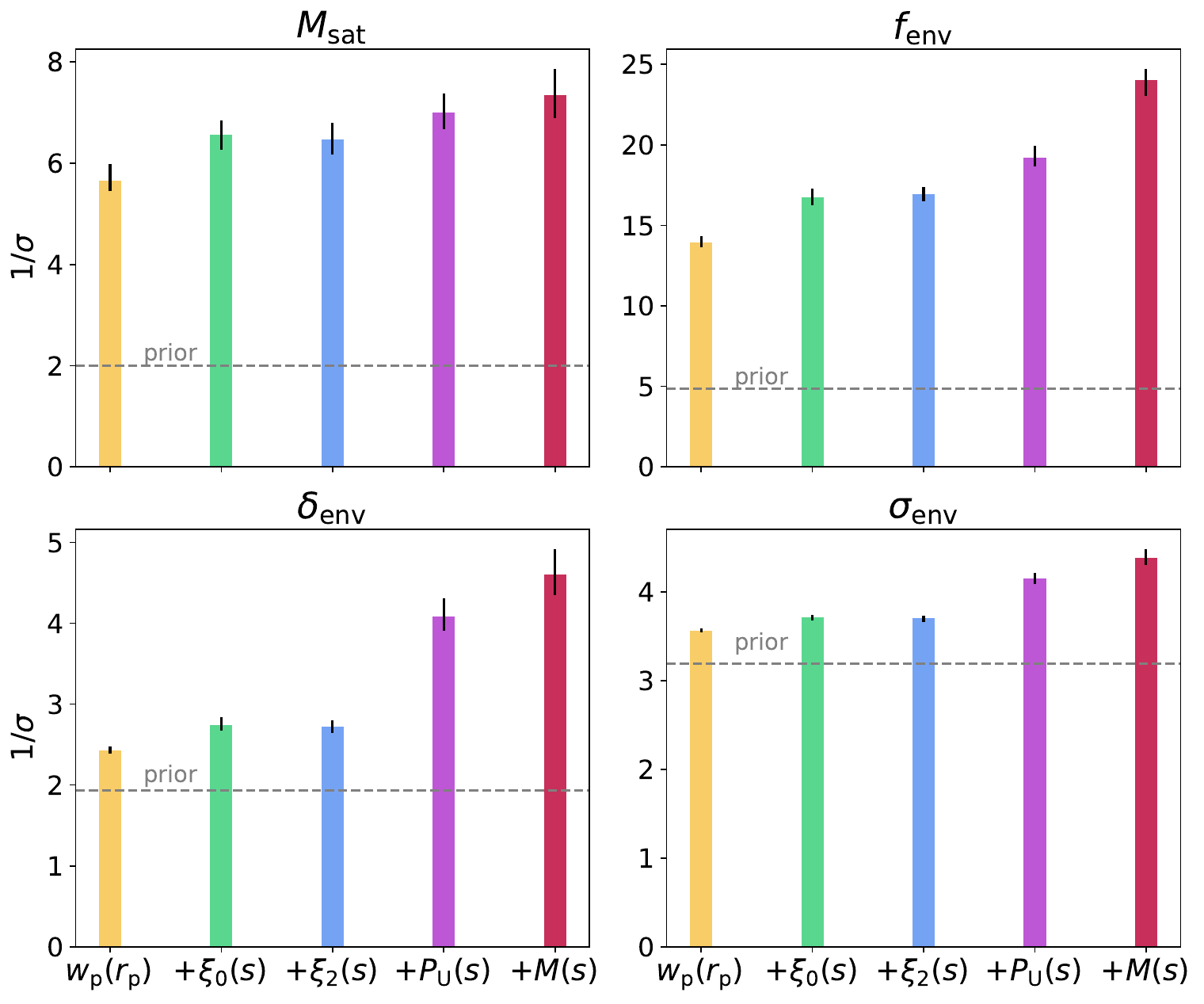}
\caption{The precision of recovery tests when successively adding in observables, averaged over the 70 test models, for the HOD parameter $\msat$ and the three assembly bias parameters. Definitions are the same as in Figure~\ref{fig:recovery_single}.}
\label{fig:recovery_hodab}
\end{figure*}

We show the precision of our recovery tests for the HOD parameter $\msat$ and the three assembly bias parameters, $\fenv$, $\delta_\mathrm{env}$, and $\sigma_\mathrm{env}$, in Figure~\ref{fig:recovery_hodab}.
Results are shown averaged over the 70 test models, when successively adding in each of our five observables.
We see that, for all of the parameters, each of the observables provides additional information on the parameter, with the exception of $\cfq$.
The two beyond-standard statistics $\upf$ and $\mcf$ provide significantly increased precision compared to the standard statistics alone.
This indicates that the additional constraining power from these statistics for the cosmological parameters may be related to their heightened sensitivity to assembly bias, as well as the ability of the combination of many observables to constrain the flexible HOD model.

It is somewhat surprising that $\wprp$ on its own provides significant constraining power over the prior on $\fenv$, the amplitude of environmental assembly bias.
Investigating the relationship between these, we find that, with the rest of the parameters fixed, at large scales, $\wprp$ decreases as $\fenv$ is increased. 
This makes sense because positive $\fenv$ values effectively transfer halos from high- to low-density regions, reducing overall clustering, which translates to a lower two-halo term.
It is notable that this effect is significant enough to be able to constrain this parameter, highlighting the importance of including a flexible model of environmental assembly bias.

\section{Full posterior plots}

\afterpage{
\begin{figure*}[p!]
\centering
\subfloat[\label{fig:contour_addin_allcosmo}]{\includegraphics[width=0.55\textwidth]{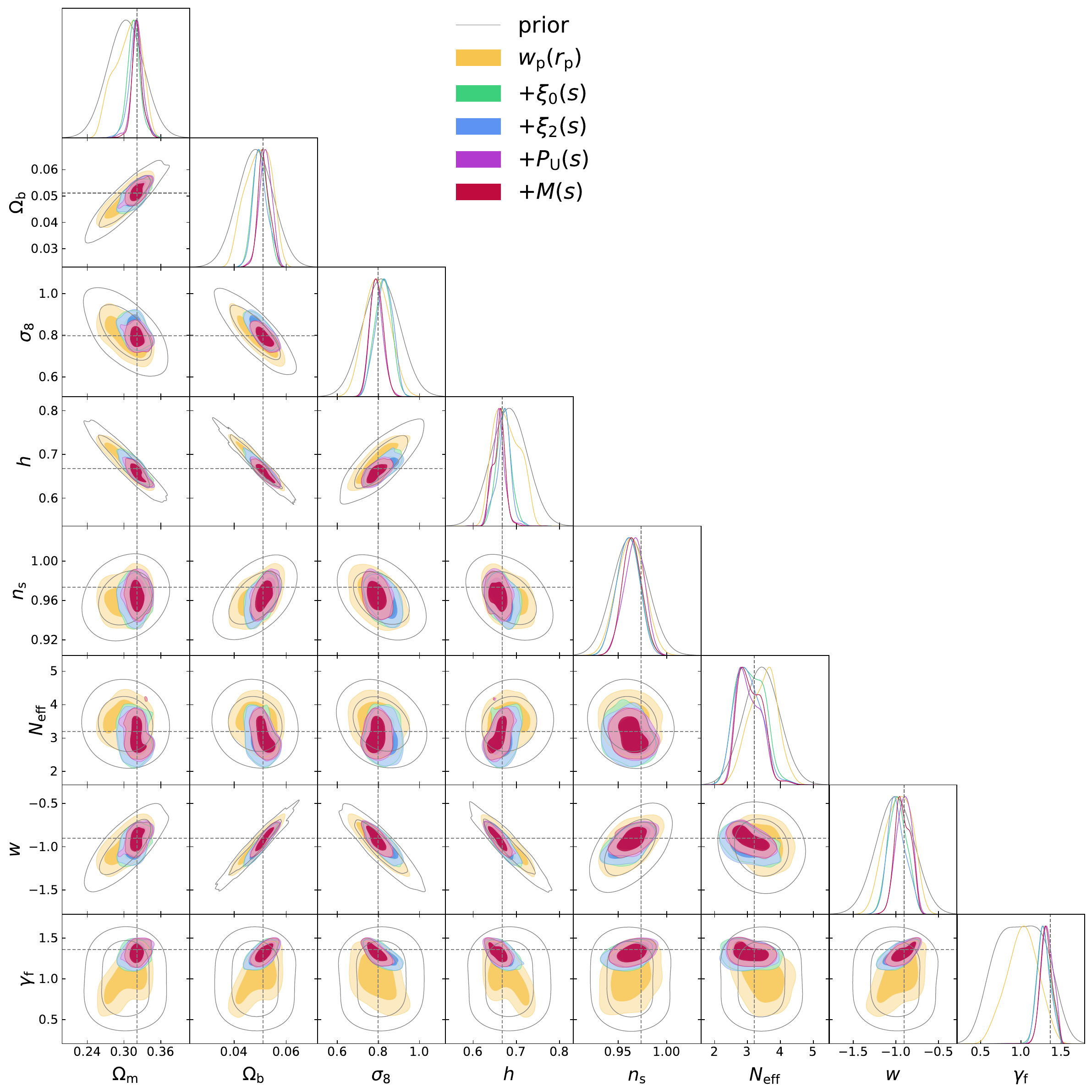}}
\vspace{1em}
\subfloat[\label{fig:contour_addin_keymix}]{\includegraphics[width=0.45\textwidth]{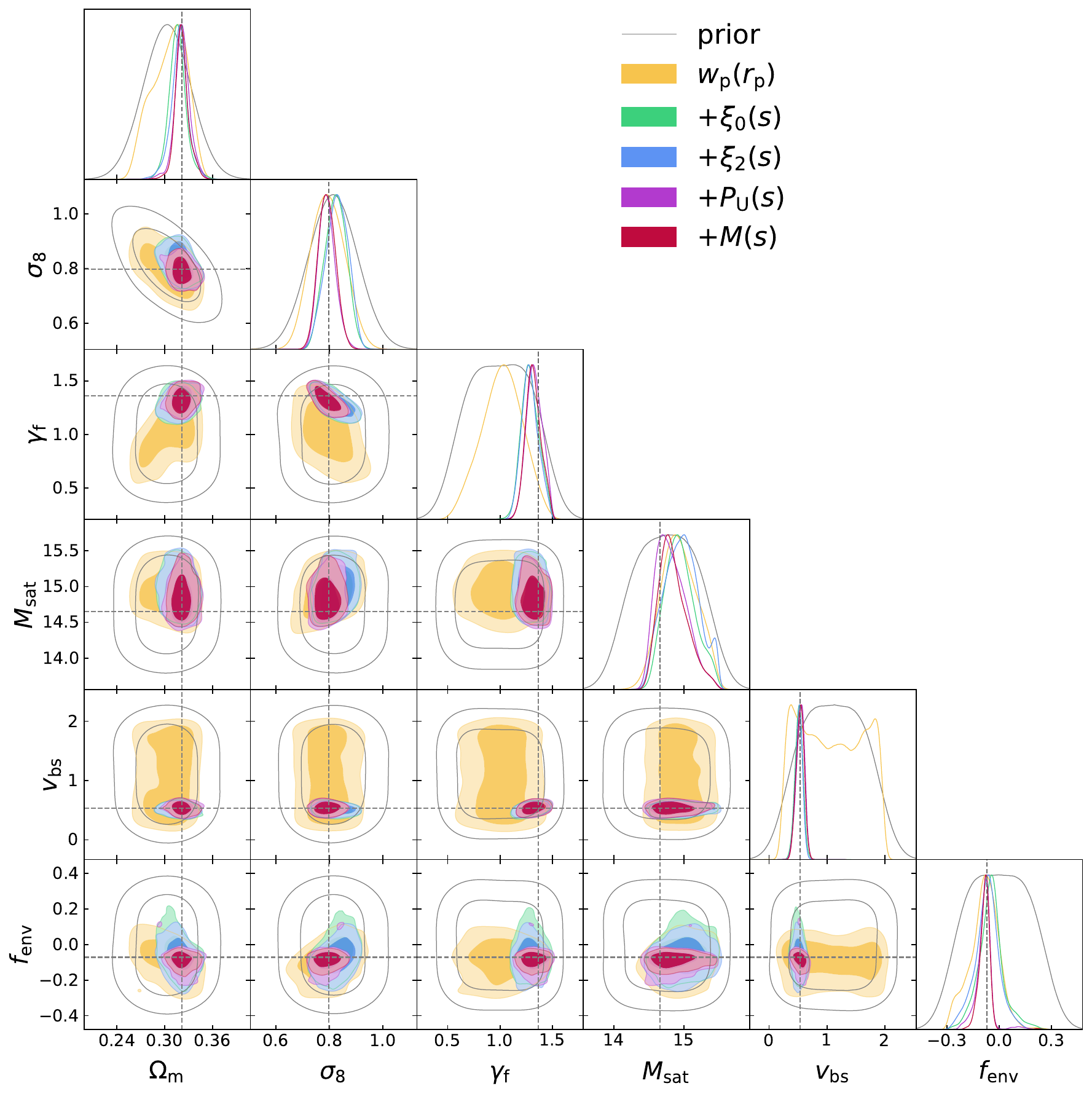}}
\caption{Posteriors for all free parameters in our recovery test of a single cosmology+HOD model, when adding in observables successively. Contours are shown for (a) all cosmological parameters; (b) a mix of the key cosmological, HOD, and assembly bias parameters; and (c) all HOD and assembly bias parameters.}
\end{figure*}
\clearpage
}

\afterpage{
\begin{figure*}[p!]\ContinuedFloat
\centering
\subfloat[\label{fig:contour_addin_allhodab}]{\includegraphics[width=0.8\textwidth]{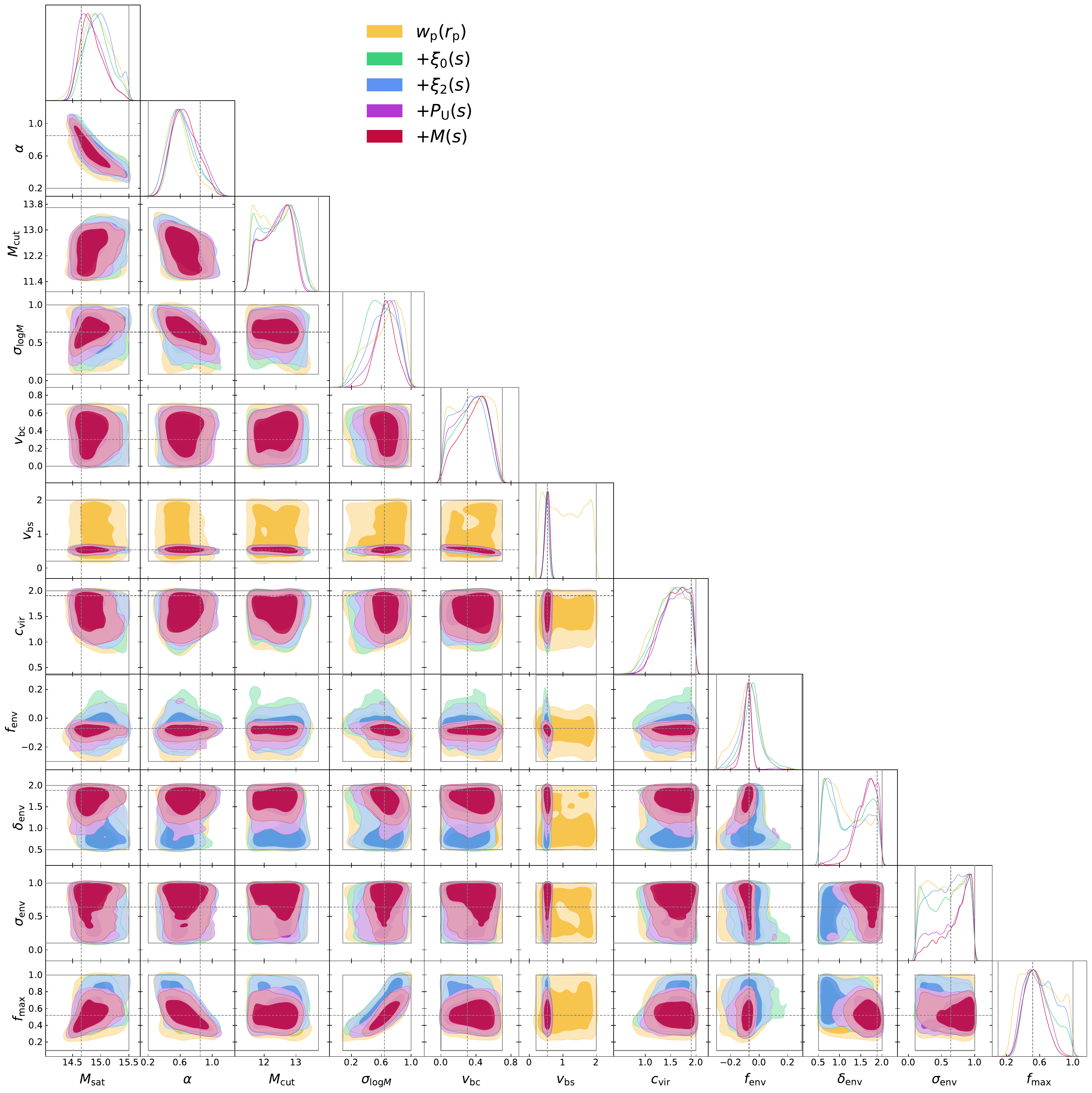}}
\caption{Continued from previous page.}
\label{fig:contours_full}
\end{figure*}
\clearpage
}

We show contour plots of all the recovered parameters for a single cosmology+HOD test model in Figure~\ref{fig:contours_full}, when successively adding in our observables.
In Figure~\ref{fig:contour_addin_allcosmo}, we show the cosmological parameters; in Figure~\ref{fig:contour_addin_keymix}, a combination of the key cosmological, HOD, and assembly bias parameters; and in Figure~\ref{fig:contour_addin_allhodab}, all the HOD and assembly bias parameters.
We can clearly see the degeneracies between many of the parameters here, and for many of these, including the beyond-standard statistics breaks the degeneracy. 
This is true for degeneracies between cosmological parameters and HOD parameters, as with $\sig$ and $\msat$; between HOD parameters, as with $\vbs$ and $\sigma_{\mathrm{log}M}$; and between assembly bias parameters, as with $\fenv$ and $\sigma_\mathrm{env}$.
This helps explain how the combination of our flexible assembly bias model and the emulation of beyond-standard statistics improves our precision on cosmological parameter constraints.

\bibliography{references}

\end{document}